\documentclass[fleqn,usenatbib]{mnras}  

\usepackage{hyperref}
\usepackage{subfigure}	
\usepackage{bm}
\usepackage{ulem}
\usepackage{color}
\usepackage{comment}
\usepackage{ae,aecompl}

\usepackage{newtxtext,newtxmath} 
\usepackage{cite}
\usepackage{enumerate}

\usepackage[T1]{fontenc} 

\DeclareRobustCommand{\VAN}[3]{#2} 
\let\VANthebibliography\thebibliography 
\def\thebibliography{\DeclareRobustCommand{\VAN}[3]{##3}\VANthebibliography}

\usepackage{graphicx}	
\usepackage{amsmath}	
\usepackage{amssymb}	
\usepackage{array}      
\usepackage{booktabs}   

\title[The Predicted IRX Relation in \textsc{EAGLE}]{The Dust Attenuation Scaling Relation of Star-Forming Galaxies in the \textsc{EAGLE} Simulations}

\author[M.~Qiao et al.]{
Man~Qiao, $^{1,2}$ Xian~Zhong~Zheng, $^{1,2}\thanks{E-mail: xzzheng@pmo.ac.cn (XZZ)}$
Antonios~Katsianis, $^{3}\thanks{E-mail: katsianis@mail.sysu.edu.cn (AK)}$ Jianbo~Qin, $^{1}$ Zhizheng~Pan,$^1$
\and
Wenhao~Liu, $^1$ Qing-Hua Tan, $^1$ Fang~Xia~An, $^1$ Dong~Dong~Shi,$^1$  Zongfei~L\"{u}, $^{1,2}$ Yuheng~Zhang, $^{1,2}$ 
\and
Run~Wen, $^{1,2}$ Shuang~Liu, $^{1,2}$ Chao~Yang $^{1,2}$
\\
$^1$Purple Mountain Observatory, Chinese Academy of Sciences, 10 Yuanhua Road, Nanjing 210023, China\\
$^2$School of Astronomy and Space Sciences, University of Science and Technology of China, Hefei 230026, China\\
$^3$School of Physics and Astronomy, Sun Yat-sen University, Zhuhai Campus, 2 Daxue Road, Xiangzhou District, Zhuhai, China \\
}
\date{Accepted 2024 January 05. Received 2024 January 04; in original form 2023 August 11}

\pubyear{2024}

\begin{document}\label{firstpage}
\pagerange{\pageref{firstpage}--\pageref{lastpage}}
\maketitle 

\begin{abstract}

Dust attenuation in star-forming galaxies (SFGs), as parameterized by the infrared excess (IRX $ \equiv L_{\rm IR}/L_{\rm UV}$), is found to be tightly correlated with star formation rate (SFR), metallicity and galaxy size, following a universal IRX relation up to $z=3$. This scaling relation can provide a fundamental constraint for theoretical models to reconcile galaxy star formation, chemical enrichment, and structural evolution across cosmic time. We attempt to reproduce the universal IRX relation over $0.1\leq z\leq 2.5$ using the EAGLE hydrodynamical simulations and examine sensitive parameters in determining galaxy dust attenuation. Our findings show that while the predicted universal IRX relation from EAGLE approximately aligns with observations at $z\leq 0.5$, noticeable disparities arise at different stellar masses and higher redshifts. Specifically, we investigate how modifying various galaxy parameters can affect the predicted universal IRX relation in comparison to the observed data. We demonstrate that the simulated gas-phase metallicity is the critical quantity for the shape of the predicted universal IRX relation. We find that the influence of the infrared luminosity and infrared excess is less important while galaxy size has virtually no significant effect. Overall, the EAGLE simulations are not able to replicate some of the observed characteristics between IRX and galaxy parameters of SFGs, emphasizing the need for further investigation and testing for our current state-of-the-art theoretical models.

\end{abstract} 




\section{INTRODUCTION}\label{sec:sec1}

It is widely accepted that galaxy formation and evolution are governed by the hierarchical assembly of dark matter haloes and baryonic physics \citep{vandenBosch2002, Mo2010, Katsianis2023}. Baryonic physics involve complex processes which shape the gas, metals and dust within the interstellar medium (ISM), the underlying stellar populations, star formation, and stellar feedback \citep{Tumlinson2017, Peroux2020, Tejos2021, Saintonge2022}. Despite the fact that dust constitutes only $\sim$1\,per\,cent of the ISM in the star-forming galaxies (SFGs), it significantly alters the panchromatic spectral energy distribution since it absorbs the ultraviolet (UV)-optical light and then re-emits it in the infrared (IR) \citep{Galliano2018, Hahn2022}. The above process seriously affects the measurements of galaxy observables \citep{Calzetti2001, Conroy2013, Trayford2020}. In addition, dust is a crucial component that drives the cycle of baryons as it facilitates the formation of molecular gas and acts as an essential coolant for star and planet formation. Furthermore, dust plays an important role in converting radiation pressure into mechanical energy of outflows \citep{Thompson2015, Naddaf2022}. Thus, a deep understanding of dust evolution and its relevant physical processes are crucial for the buildup of a comprehensive picture of galaxy evolution.

Although our current state-of-the-art models are able to globally predict the dust content for galaxy populations up to redshift $z=9$ \citep{Popping2017, Dave2019}, there is a major challenge in simultaneously reproducing the observed fundamental properties of galaxies (e.g. the star formation rate function, the cosmic star formation rate density, the H\,{\small I} mass function, and the H\,{\small I}-to-halo mass relation) at different epochs in a self-consistent framework \citep{Zhao2020, Katsianis2021, Li2022, Picouet2023}. The dust content of a galaxy is determined by the amount of gas and the gas-to-dust ratio, which is controlled by the gas-phase metallicity \citep{Trayford2017, Galliano2018, DeVis2019, Trcka2020, Casasola2020}. Observationally, the attenuated stellar emission and dust thermal IR radiation of galaxies can be directly measured while the dust content is indirectly inferred and largely model dependent. Dust attenuation (or obscuration) describes the extent at which the stellar radiation in a galaxy is absorbed and scattered away from and back to the line-of-sight by dust \citep{Salim2020}. In practice, a luminosity ratio between the IR to the UV (i.e. infrared excess IRX $ \equiv L_{\rm IR}/L_{\rm UV}$) is often used to measure dust attenuation \citep{Heckman1998, Meurer1999, Heinis2013, Bourne2017}. The strength of dust attenuation is commonly regulated by the star formation density and the effective dust column density, and thus is tightly coupled with star formation, metallicity, and galaxy compactness \citep{Garn2010, Xiao2012, Battisti2016, Katsianis2017a, Zahid2017, Shapley2022}.

Star formation, chemical enrichment, and structural growth are three key factors governing galaxy evolution. Star formation in SFGs is driven by the gas content and the pre-existing stars, following the Kennicutt-Schmidt Law ($\Sigma_{\rm gas}$--$\Sigma_{\rm SFR}$) \citep{Kennicutt1998} and the stellar mass--star formation rate ($M_\ast$--SFR) relation \citep{Matthee2019, Katsianis2020, Thorne2021, Leja2022, Kouroumpatzakis2023}. The abundance of metals in the ISM is correlated with stellar mass and forms the stellar mass--metallicity ($M_\ast$--$Z$) relation \citep{Blanc2019, Bellstedt2021, Lewis2023}. The size (i.e. half-light radius $R_{\rm e}$ in the rest-frame optical band) of disc-dominated SFGs increases with increasing stellar mass, giving the stellar mass--size ($M_\ast$--$R_{\rm e}$) relation \citep{Ichikawa2012, Nedkova2021, Yang2021}. These scaling relations are well established out to $z\sim 3$ and demonstrate rapid evolution over this redshift range.

Dust attenuation IRX is found to increase with stellar mass for SFGs with $M_{\ast} <10^{10.5}\,{\rm M_\odot}$, however, little change is observed in this relation at $z<3$ \citep{Whitaker2014, Heinis2014, Bouwens2016, McLure2018, Shapley2022}. From the evolving $M_{\ast}$--SFR, $M_{\ast}$--$Z$, and $M_{\ast}$--$R_{\rm e}$ relations, combined with the Kennicutt-Schmidt Law and a constant metal-to-dust ratio, one can infer a significant increase in dust mass surface density ($\Sigma_{\rm dust}$) with increasing redshift \citep{Shibuya2015, Wuyts2016}, and thus a higher IRX for high-$z$ SFGs at a fixed stellar mass, which contradicts observations (see \citealt{Wuyts2011}, \citealt{Genzel2013} and \citealt{Qin2019} for more details).

Interestingly, \citet{Qin2019} discovered that the observed IRX is determined by $Z$, IR luminosity ($L_{\rm IR}$), $R_{\rm e}$ and galaxy inclination ($b/a$) in a form of joint power-law function as ${\rm IRX} = 10^\alpha\,L_{\rm IR}^\beta\,R_{\rm e}^{-\gamma}\,{(b/a)}^{-\delta}$, with these power-law indices all a function of gas-phase metallicity, giving a universal relation for SFGs up to $z=3$. This universal IRX relation unifies the existing correlations between dust attenuation and other galaxy parameters, and describes the relative contributions by metallicity, star formation ($L_{\rm IR}$), galaxy compactness ($R_{\rm e}$), and galaxy inclination. Most importantly, the gas-phase metallicity was found to play a central role in regulating the IRX through a manner that was totally unexpected before. As gas-phase metallicity decreases, all power-law indices decrease, and IRX declines in a way totally different from the estimate based on the Kennicutt-Schmidt Law and a constant metal-to-dust ratio \citep{Wuyts2011, Genzel2013, Qin2019}. This discovery demonstrates that for the SFGs with $M_{\ast} <10^{10.5}\,{\rm M_\odot}$ showing ``no evolution on IRX'' over $0<z<3$ is no longer puzzling. The universal IRX relation reflects that star formation, ISM metallicity, and galaxy structure jointly determine dust attenuation in SFGs, making it a fundamental relationship to test theoretical models.

We use the state-of-the-art Evolution and Assembly of GaLaxies and their Environments (EAGLE) simulations to investigate the theoretically predicted IRX in relation with other galaxy parameters, and see if the observed universal IRX relation can be recovered from the EAGLE simulations. The cosmological hydrodynamical EAGLE simulations are a valuable tool for understanding the physical mechanisms and determinants of dust attenuation while the universal IRX relation is closely linked with the baryon cycles occurring in galaxies. Thus, the comparison between the two can in turn highlight successes of the simulations and identify areas for further improvement.

This paper is organized as follows. We introduce the EAGLE simulations, sample selection, and galaxy parameter estimation in Section~\ref{sec:sec2}. In Section~\ref{sec:sec3}, the predicted universal IRX relation is presented, and we compare it with the observations, and focus on the galaxy parameters that should be improved in future efforts of state-of-the-art simulations in order to describe better the observations. We discuss our results in Section~\ref{sec:sec4} and give a summary in Section~\ref{sec:sec5}.

\begin{table*} 
    \centering 
    \caption{Simulation parameters of the EAGLE reference model Ref-L0100N1504. Columns from left to right refer to model name, comoving box size, number of dark matter/baryonic particles, initial baryonic particle mass, dark matter particle mass, comoving Plummer-equivalent gravitational softening length, maximum proper softening length, the number of galaxies with $M_{\ast} > 10^{8.5}$\,M$_{\odot}$, the number of galaxies with $N_{\rm dust} > 0$ (``some dust'') and with $N_{\rm dust}$ > 250 (``resolved dust'').} 
    \label{tab:tab1}
    \begin{tabular} {m{2cm}<{\centering} m{0.7cm}<{\centering} m{0.7cm}<{\centering} m{1.2cm}<{\centering} m{1.2cm}<{\centering} m{0.7cm}<{\centering} m{0.7cm}<{\centering} m{0.9cm}<{\centering} m{2.1cm}<{\centering} m{2.1cm}<{\centering}} 
        \toprule 
        Name & $L$ & $N$ & $m_{\rm g}$ & $m_{\rm dm}$ & $\epsilon_{\rm com}$ & $\epsilon_{\rm prop}$ & \multicolumn{3}{c}{Number of galaxies with $M_{\ast}$ > $10^{8.5}$ $\rm {M_\odot}$}\\ 
        \cmidrule(lr){8-10} 
        & (cMpc) & & ($\rm {M_\odot}$) & ($\rm {M_\odot}$) & (ckpc) & (pkpc) & All & With some dust & With resolved dust\\
        \midrule 
        Ref-L0100N1504 & 100 & $1504^3$ & $1.81\times 10^6$ & $9.70\times 10^6$ & 2.66 & 0.70 & 371,728 & 334,717 (90.0\%) & 236,346 (63.6\%)\\
        \bottomrule 
    \end{tabular}
\end{table*}

\section{SIMULATION and DATA}\label{sec:sec2}

\subsection{The EAGLE Simulations}\label{sec:sec2.1}

The EAGLE project \citep{Crain2015, Schaye2015} was a suite of cosmological, hydrodynamical simulations of galaxy formation within the cubic volumes ranging from 25 to 100 co-moving Mpc (cMpc). As one of the state-of-the-art cosmological, hydrodynamical simulations in the framework of the standard $\Lambda$ cold dark matter cosmology, the EAGLE simulations modeled the evolution of baryonic matter and cold dark matter from redshift $z=127$ to the present day in a self-consistent manner. The EAGLE simulations made use of an improved version of the N-Body Tree-PM smoothed particle hydrodynamics (SPH) code \textsc{GADGET-3} (the readers are referred to \citealt{Springel2005} for more details), which included a modified hydrodynamics solver, time-step limiters \citep{Durier2012}, and sub-grid treatments of baryonic physics. A Chabrier stellar initial mass function (IMF) \citep{Chabrier2003} was adopted. The EAGLE simulations utilized sub-grid models for computing radiative cooling, star formation, stellar mass loss \citep{Wiersma2009}, metal enrichment, stellar and active galactic nuclei (AGN) feedback, gas accretion onto supermassive black hole (SMBH), and SMBH mergers.

The EAGLE simulations were calibrated to the redshift $z=0$ galaxy stellar mass function, galaxy stellar mass--size relation, and stellar mass--SMBH mass relation. A large number of studies based on the EAGLE simulations investigated the extent to which they reproduce those different observed variables that are not used for their calibrations \citep[e.g.][]{Furlong2015, Camps2016, Furlong2017, Katsianis2017b, Lovell2022, Ward2022, Das2023, Keller2023, Petit2023}. The SUBFIND algorithm \citep{Springel2001, Dolag2009} was employed to identify gravitationally-bound sub-haloes as individual galaxies, and the sum of the mass of all star particles that belong to a sub-halo was determined the galaxy stellar mass. The EAGLE collaboration provided a comprehensive dataset for galaxies, spanning 29 redshifts over $0<z<20$ \citep{McAlpine2016}. Additionally, star formation was set to follow the pressure-dependent Kennicutt-Schmidt relation \citep{Schaye2008}, with a metallicity-dependent density threshold adopted to track the transition of warm and atomic gas into unresolved, cold, molecular gas \citep{Schaye2004}. Furthermore, the EAGLE simulations tracked individually the abundance of 11 elements (H, He, C, N, O, Ne, Mg, Si, S, Ca and Fe; \citealt{DeRossi2017}).

Dust attenuation and dust emission for sufficiently resolved galaxies with stellar mass $>10^{8.5}$\,$\rm {M_\odot}$ were calculated by \citet{Camps2018} using the three-dimensional dust radiative transfer code \textsc{SKIRT} \citep{Baes2011, Camps2015}, in which the \citet{Zubko2004} dust model was adopted to represent the diffuse dust, and a similar dust model was used for the star-forming regions following the MAPPINGS-III templates \citep{Groves2008}. The dust grains in these models include a mixture of non-composite graphite and silicate grains, and neutral and ionized polycyclic aromatic hydrocarbon (PAH) molecules. A fixed dust-to-metal ratio of 0.3 was used regardless of galaxy type or redshift \citep{Camps2018}. Stochastic heating of dust grains was taken into account in SKIRT to obtain realistic 50-band fluxes across the UV to submillimetre spectral range \citep{Trayford2017, Baes2019, Katsianis2019, deGraaff2022, Trcka2020, Evans2022, Cochrane2023}. We refer the readers to \citet{Camps2016}, \citet{Trayford2017}, and \citet{Camps2018} for more details.

\subsection{Sample Selection}\label{sec:sec2.2}

\begin{figure*} 
	\centering
	\includegraphics[width=0.9\textwidth]{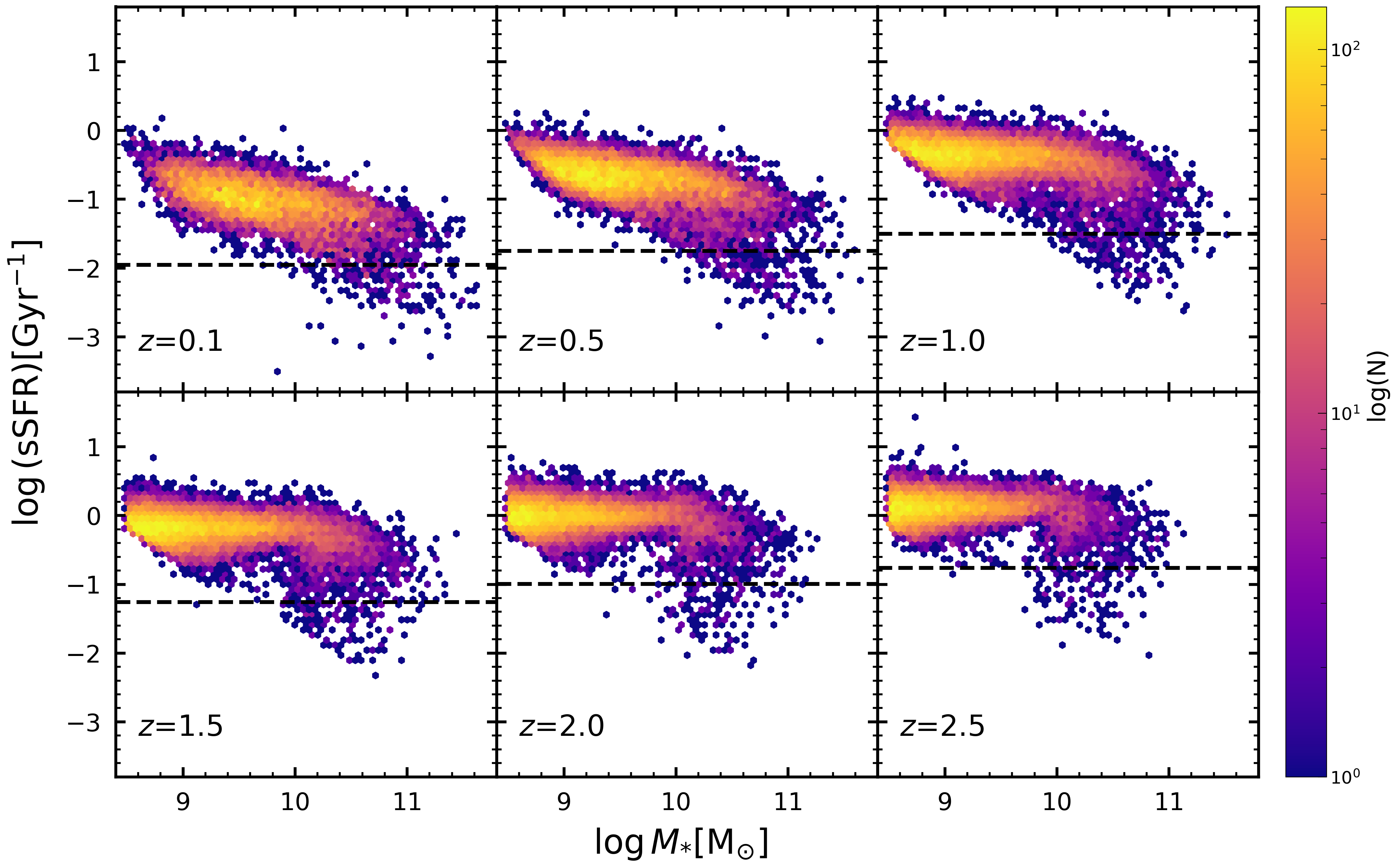}
	\caption{The $M_{\ast}$--sSFR relations of simulated galaxies from the EAGLE Ref-L0100N1504 model at six redshifts over $0.1\leq z \leq 2.5$. The colour of the hexagonal bins represents the number of sample galaxies within that bin. The black dotted line refers to the sSFR cut to separate SFGs and QGs at each redshift. The number of SFGs above the black dotted line in each panel from $z=0.1$ to $z=2.5$ is 7867, 11609, 15867, 16835, 15070 and 12187, respectively.}\label{fig:fig1} 
\end{figure*}

We make use of the publicly available data of the EAGLE reference model Ref-L0100N1504.\footnote{http://www.eaglesim.org/database.php} Note that adoption of the EAGLE high-resolution model Recal-L025N0752 with a smaller sample of galaxies will give quantitatively consistent results. Table~\ref{tab:tab1} summarises the key parameters of the Ref-L0100N1504 model for 371,728 galaxies with stellar mass above $10^{8.5}$\,M$_{\odot}$. This stellar mass cutoff matches the set of galaxies for which the public EAGLE database includes optical magnitudes without dust attenuation. The fluxes from SKIRT are meaningful only if the input distributions for both the stellar sources and the dust content are spatially resolved to an acceptable level. The model used a number of smoothed (sub)particles ($N_{\rm dust}$) to count dust content, and we adopt $N_{\rm dust}>250$ to recognize sample galaxies with resolved dust density distribution (see \citealt{Camps2018} for more details). We follow the description in \citet{McAlpine2016} for the use of the EAGLE dataset to select our sample galaxies with parameters of their intrinsic properties, and adopt dust-attenuated fluxes and dust emission fluxes from \citet{Camps2018}. We select galaxies from six snapshots, corresponding redshifts $z$ = 0.1, 0.5, 1.0, 1.5, 2.0 and 2.5, respectively. The galaxy parameters, including stellar mass, star formation rate, half-mass radius ($R_{\rm e,mass}$), absolute abundances of hydrogen and oxygen of star-forming gas, and broadband fluxes, are all derived from a spherical aperture of 30\,proper\,kpc (pkpc). The UV/optical-band fluxes are included for galaxies viewed at face-on, edge-on and random  angles \citep{Camps2018}. We choose the fluxes for galaxies at the random viewing angles, corresponding to the random orientation of galaxies in observations.

We separate the SFGs from quiescent galaxies (QGs) in the six snapshots of the Ref-L0100N1504 model, using a redshift dependent specific star formation rate ($\rm sSFR$ $\equiv$ SFR/$M_{\ast}$) cut, described as $\rm \log_{10}\big(sSFR_{\lim}$ ($z$)$\rm /Gyr^{-1}\big)$ = 0.5$z$ $-$ 2 with $z \in [0, 2]$ following \citet{Furlong2017}. We extrapolate the cut to $z=2.5$ to fit the redshift range for our sample. Note that at $z\geq1$ a subsample of low-mass EAGLE SFGs ($M_\ast \lesssim 10^{9.5}$\,M$_\odot$) exhibit compact ($R_{\rm e}<0.5$\,kpc), metal-rich ($12+\log{\rm (O/H)} \gtrsim 9.0$) and dusty (IRX\,$\gtrsim 1$) characteristics. These populations resemble spurious galaxies \citep{McAlpine2016}, and we remove them from our sample. Fig.~\ref{fig:fig1} shows the $M_{\ast}-$sSFR relation in six redshifts, $z =$ 0.1, 0.5, 1.0, 1.5, 2.0 and 2.5. We note that the simulated galaxies from the EAGLE Ref-L0100N1504 model are mostly SFGs, and the bimodality is not so prominent in the sSFR compared to the observations \citep{Furlong2015, Trcka2020}. This limitation is prevalent for various models including TNG \citep{Katsianis2021b}. However, this limitation does not affect the results of our work, which focuses on SFGs. As expected at increasing redshift, the number of massive SFGs ($M_{\ast}$ > $10^{10}$\,$\rm {M_\odot}$) decreases, while the average sSFR gradually increases. In contrast, the low-mass galaxies ($M_{\ast}$ < $10^9$\,$\rm {M_\odot}$) are numerous at all redshifts. Observationally, these low-mass galaxies are hardly detected at higher redshift and thus less explored.

\begin{figure*}
	\centering
	\includegraphics[width=0.9\textwidth]{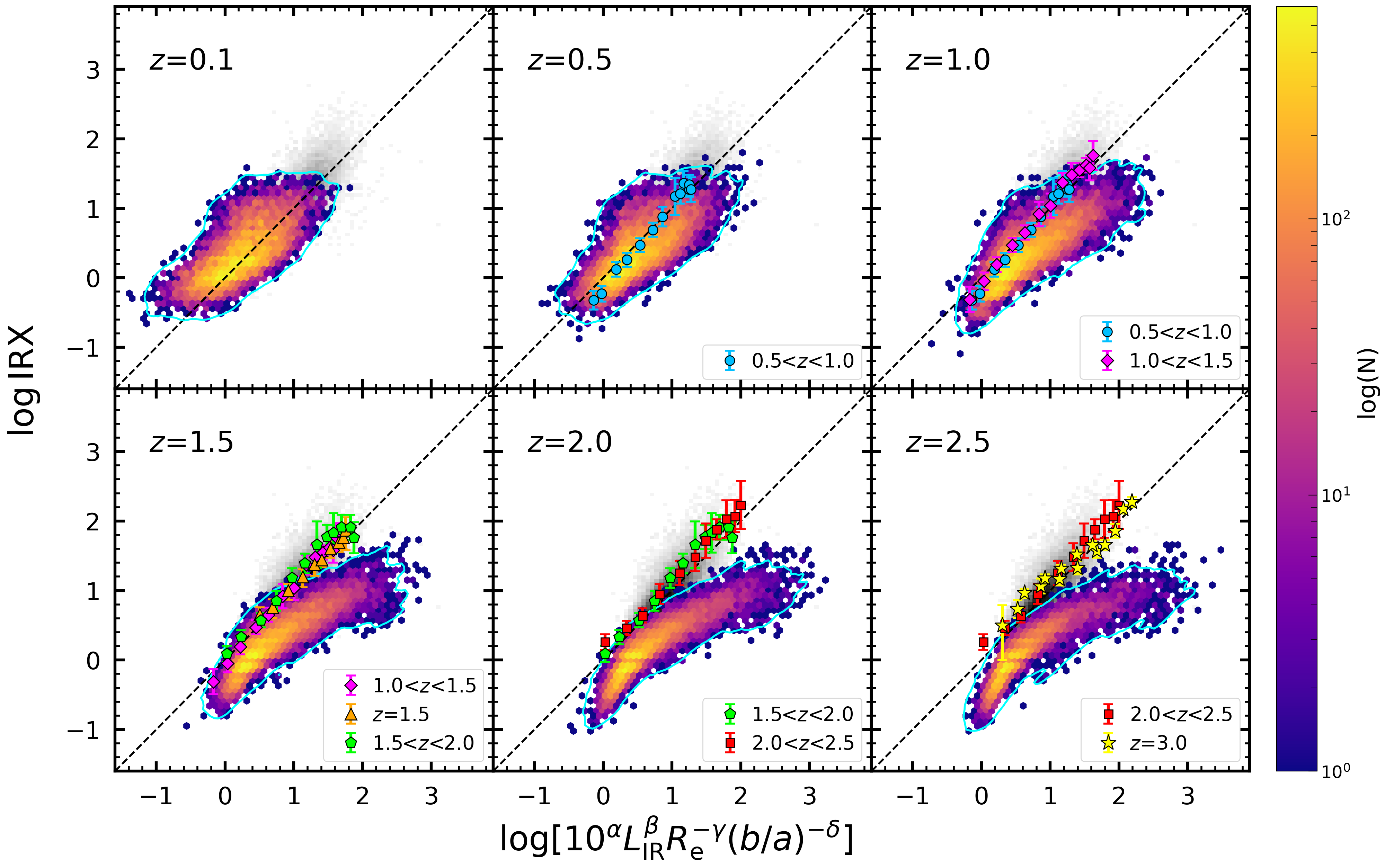}
    \caption{Distributions of our EAGLE SFGs (coloured hexagonal bins) at six redshifts compared with local SFGs (gray-scale image) from \protect\citet{Qin2019} in the universal IRX relation. The data points with error bars refer to the average SFGs of different stellar masses at given redshifts from \protect\citet{Whitaker2014} for $0.5<z<2.5$, \protect\citet{Heinis2014} for $z=1.5$ and $z=3.0$, and \protect\citet{AlvarezMarquez2016} for $z=3$. The colour of the hexagonal bin represents the number of SFGs within that bin. The cyan contour marks the outer outline of the distribution of SFGs. One can see that the EAGLE SFGs roughly follow the observed relation, while noticeable deviation is visible, especially for the EAGLE SFGs at $z\geq 1$.}\label{fig:fig2}
\end{figure*}

\subsection{Galaxy Parameters}\label{sec:sec2.3}

\subsubsection{Gas-Phase Metallicity}\label{sec:sec2.3.1}

We obtain oxygen abundance $12+\log {\rm (O/H)}$ from the masses of oxygen and hydrogen in the star-forming gas as the gas-phase metallicity for our EAGLE SFGs. The oxygen abundance is therefore the absolute abundance which does not depend on the solar abundance \citep{Camps2018}.

\subsubsection{Half-Light Radius}\label{sec:sec2.3.2}

We focus on the parameters employed to construct the universal IRX relation from \citet{Qin2019}. The half-mass radius $R_{\rm e,mass}$ from the simulation catalogue can be converted into half-light radius $R_{\rm e}$ which is widely used in observations. \citet{Suess2019b} studied the half-mass-to-half-light ratio ($R_{\rm e,mass}/R_{\rm e}$) for $\sim$16,500 galaxies at $0 \lesssim z \leq 2.5$ in the CANDELS fields, and found that for a fixed stellar mass, the $R_{\rm e,mass}/R_{\rm e}$ remained roughly constant at $0<z<1$, but evolved rapidly between $z\sim 1$ and $z\sim 2.5$. We read the $M_{\ast}$ versus $R_{\rm e,mass}/R_{\rm e}$ relations of SFGs in five redshift bins from their Fig. 1, and interpolate them to obtain  $R_{\rm e,mass}/R_{\rm e} = k \times M_{\ast} + b$ at six redshifts $z=[0.1, 0.5, 1.0, 1.5, 2.0, 2.5]$, yielding $k=[-0.15, -0.18, -0.21, -0.21, -0.16, -0.10]$ and $b=[2.25, 2.56, 2.94, 2.97, 2.56, 2.09]$. These relations are used to convert $R_{\rm e,mass}$ into $R_{\rm e}$ for our EAGLE SFGs. Note that we extrapolate the $M_{\ast}$--$R_{\rm e,mass}/R_{\rm e}$ relation into the low-mass regime of $10^{8.5 - 9}$\,M$_\odot$, and  the $R_{\rm e,mass}/R_{\rm e}$ ratio is set to be saturated to unity for all redshifts \citep{Suess2019a, Suess2019b, Miller2023}. \citet{deGraaff2022} constructed mock images of $z=0.1$ Ref-L0100N1504 EAGLE galaxies with SKIRT by accounting for the effects of stellar population gradients and dust, as well as instrumental effects and projection effects, and measured the half-light radius in the same way as in observational studies. The same tendency between the half-mass radius and half-light radius was found for SFGs as the increasing half-mass-to-half-light ratio at decreasing stellar mass shown in \citet{Suess2019b}, which is adopted in our work for half-light radius. We make comparison between our work and \citet{deGraaff2022} in Fig.~\ref{fig:figA2}, and find that the two are consistent with each other.

\subsubsection{UV Luminosity, IR Luminosity and IR Excess}\label{sec:sec2.3.3}

To measure the luminosity, we use the Code Investigating GALaxy Emission\footnote{https://cigale.lam.fr/} \citep[\textsc{CIGALE},][]{Boquien2019} to fit the broadband SEDs of our EAGLE SFGs, covering FUV, NUV, $u, g, r, i, z$, $J$, $H$, and $K_{\rm s}$. We use stellar spectral templates from the BC03 stellar population synthesis model \citep{Bruzual2003} with the Chabrier initial mass function \citep{Chabrier2003} and solar metallicity. A delayed star formation history additional with a recent starburst is adopted. The stellar age varies from 1 to 13\,Gyr, and the e-folding time varies from 5 to 8\,Gyr. An additional recent decline starburst is adopted. The burst age and e-folding time are in the range of  20$-$500\,Myr and 5$-$100\,Myr respectively.  For the dust attenuation, we adopt a fixed Calzetti attenuation curve \citep{Calzetti2000} with E(B$-$V) varies from 0.0 to 0.3\,mag. For each galaxy, the UV luminosity  (rest-frame 1216$-$3000\,\AA) is calculated from the integration of the best-fitting galaxy SED.

We measure the IR luminosity via best-fitting the simulated IR data points with the IR SED templates from the dust radiation model by \citet{Draine2007}. The dust radiation model covers parameter space from 0.47 to 4.58 for the PAH fraction, the minimum radiation field $U_{\rm min}=[0.1, 25]$, the maximum radiation field $U_{\rm max}=[10^3, 10^6]$, and the fraction illuminated from $U_{\rm min}$ to $U_{\rm max}$ is [0, 1]. If the IR bands which fall below 3\,$\sigma$ do have valid flux and error measurements, they are also included to constrain the IR luminosity measurements. For the bands without detections, the upper limits are used in the fitting.

Using the $L_{\rm UV}$ and $L_{\rm IR}$, we calculate the IRX for our EAGLE SFGs. Fig.~\ref{fig:figA1} shows the relationships between the IRX and the main physical quantities, including $M_{\ast}$, SFR, $Z$, $L_{\rm IR}$ and $R_{\rm e}$, for our sample of the EAGLE SFGs at six different redshifts over $0.1 \leq z \leq 2.5$. It is clear that $M_{\ast}$, SFR, $Z$ and $L_{\rm IR}$ all strongly correlate positively with IRX, whereas $R_{\rm e}$ displays a relatively negative weak correlation with IRX.

\begin{figure*}
	\centering
	\includegraphics[width=0.9\textwidth]{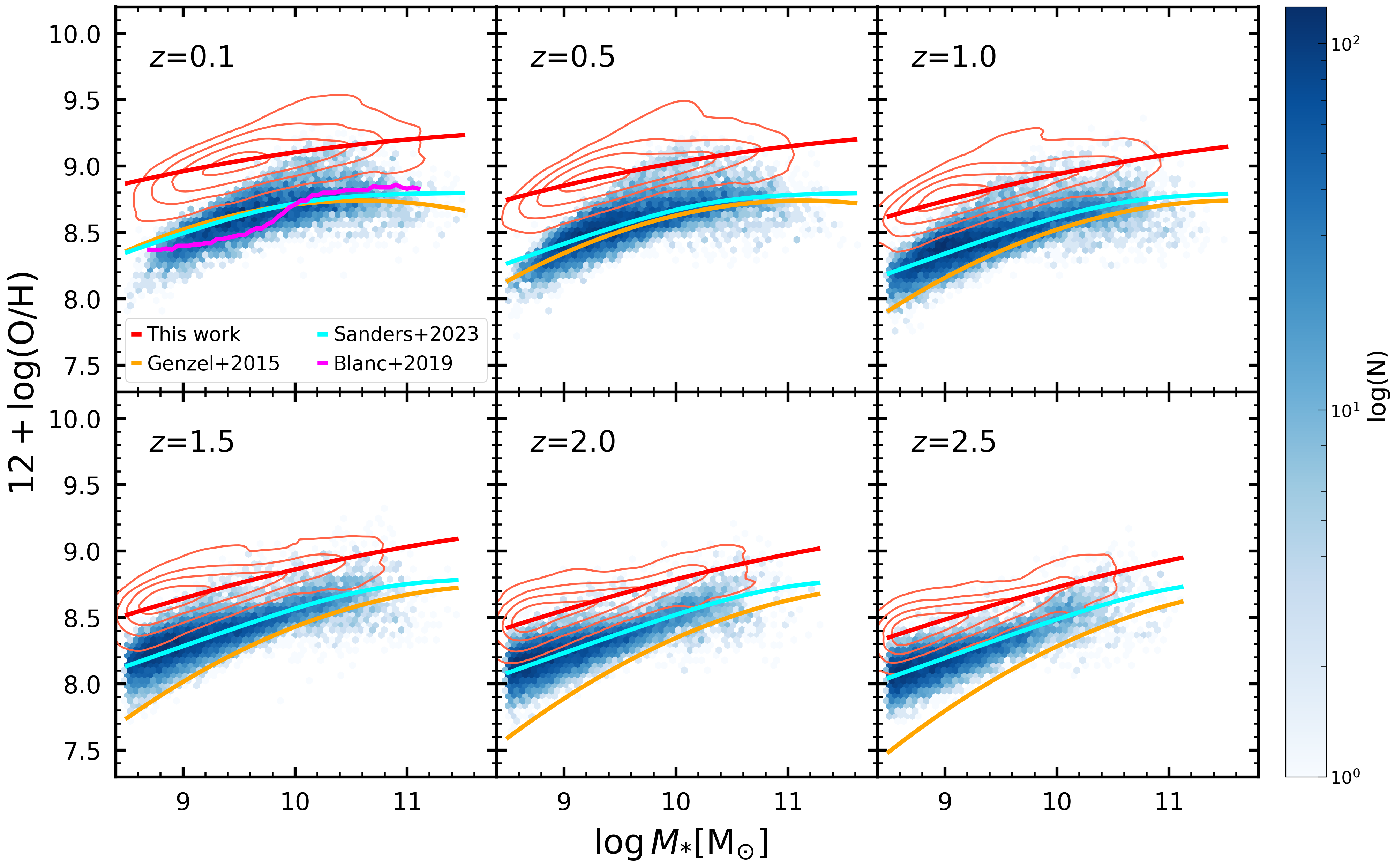}
    \caption{The stellar mass--metallicity relations of the EAGLE SFGs at six redshifts. The red contours show the distributions of the original stellar mass and metallicity, while the red lines represent the best-fitting relations in the form of Equations~\ref{eq:eq2}. The cyan lines mark the $M_{\ast}$--$Z$ relations from \protect\citet{Sanders2023}, and  the hexagonal bins are colour coded by the number of the EAGLE SFGs corrected for the metallicity offset within that bin. Clearly, the EAGLE SFGs appear to be more metal rich compared with the observed counterparts at a given stellar mass at all redshifts. The orange lines represent the $M_{\ast}$--$Z$ relations from \protect\citet{Genzel2015}, while the magenta line comes from \protect\citet{Blanc2019}.}\label{fig:fig3}
\end{figure*}

\section{RESULTS}\label{sec:sec3}

The universal IRX relation reported by \citet{Qin2019} is a joint power-law relation between $Z$, $L_{\rm IR}$, $R_{\rm e}$, $b/a$ and IRX, following 
\begin{equation}\label{eq:eq1}
\begin{split}
& IRX = 10^\alpha\,L_{\rm IR}^\beta\,R_{\rm e}^{-\gamma}\,{(b/a)}^{-\delta}, \mathrm{with}\\
& \alpha = 1.07\log (Z/Z_\odot) + 0.95, \\
& \beta = 0.91\log (Z/Z_\odot) + 0.64, \\
& \gamma = 1.15\log (Z/Z_\odot) + 0.80, \\
& \delta = 1.43\log (Z/Z_\odot) + 0.94.
\end{split}
\end{equation}
The observed IRX increases with increasing $Z$ and $L_{\rm IR}$, but decreases with increasing $R_{\rm e}$ and $b/a$. The power-law indices all exponentially increase with gas-phase metallicity, denoting that the gas-phase metallicity is a key driver of galaxy dust attenuation.

\subsection{Comparison between Simulations and Observations}\label{sec:sec3.1}

Due to the randomly distributed inclination of the EAGLE SFGs, we employ a constant $b/a$ value of 0.6 \citep{Qin2019} in the following study. Using carefully calculated $Z$, $L_{\rm IR}$, $R_{\rm e}$ and $b/a$ values for our EAGLE SFGs, we derive the predicted universal IRX relation with Equation~\ref{eq:eq1}. Fig.~\ref{fig:fig2} compares the universal IRX relation between the EAGLE simulations and the observations. In each panel, the coloured hexagonal bins represent the EAGLE SFGs in our sample, while the gray-scale image refers to the density map of local SFGs obeying the universal IRX relation from \citet{Qin2019}.

At $z\leq 0.5$, the EAGLE SFGs follow the universal IRX relation but show a relatively larger scatter. At $z\geq 1.0$, the simulated SFGs in EAGLE increasingly deviate from the relation towards the right side in each panel, particularly for the high-IRX end. For instance, the Milky-Way-like galaxies with $\log (M_\ast/$M$_\odot)\sim10.5$ at $z=2$ appear one order of magnitude lower in IRX than the inferred IRX from their parameters ($Z$, $L_{\rm IR}$, $R_{\rm e}$ and $b/a$). Similarly, the low-IRX SFGs in EAGLE also exhibit an increasing deviation from the universal IRX relation at increasing redshift. We point out that the deviation of our EAGLE SFGs from the observed universal IRX relation are clear at $z\geq 1.0$.

In the next sections we address the scaling relations of the simulated SFGs to examine what are responsible for the deviations in the predicted universal IRX relation.

\begin{figure*}
	\centering
	\includegraphics[width=0.9\textwidth]{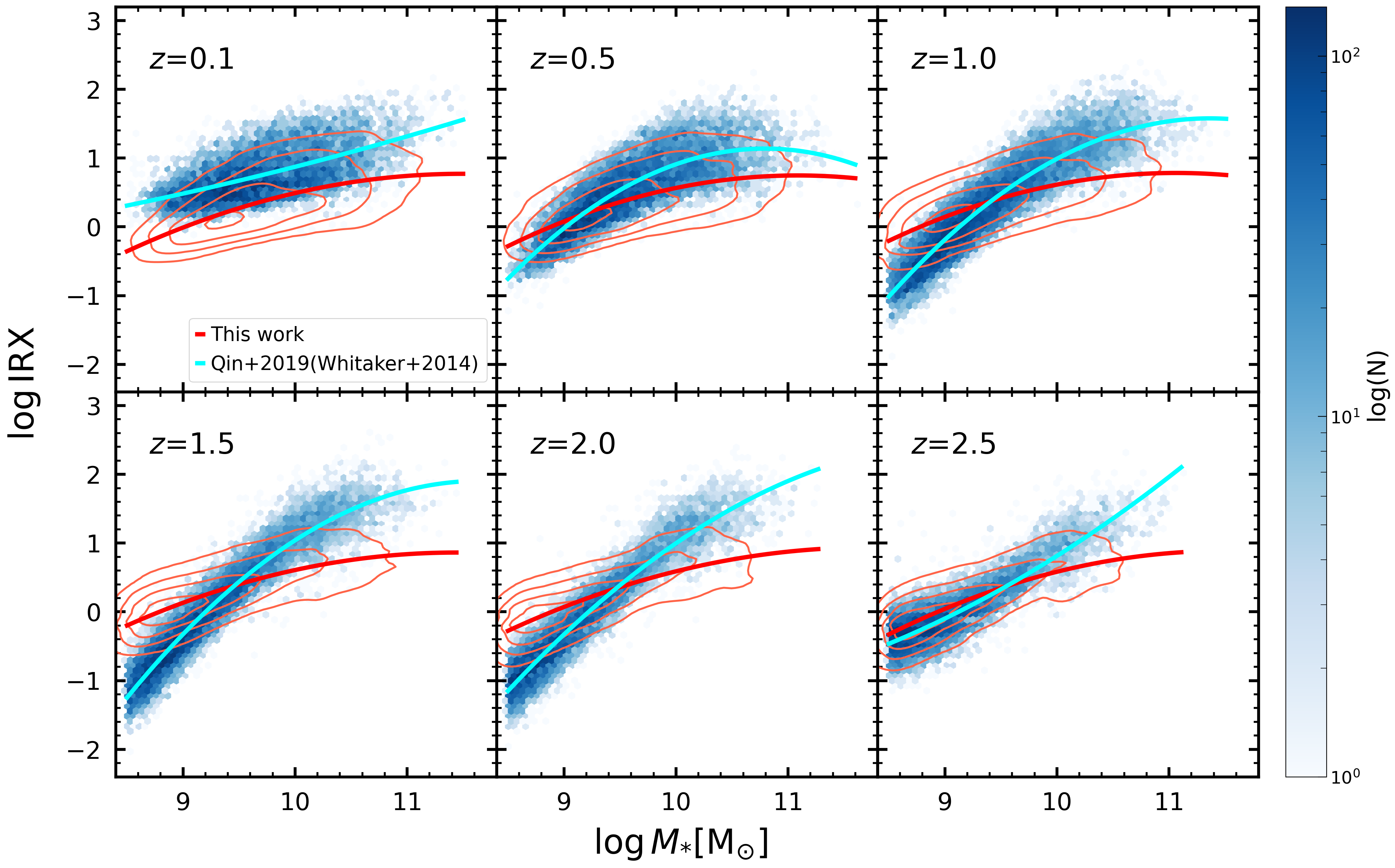}
    \caption{The stellar mass--IRX relations for our EAGLE SFGs with original (red contours) and corrected (coloured hexagonal bins) IRX  at six redshifts. The colour of the hexagonal bins represents the number of the EAGLE SFGs with corrected IRX within that bin. The cyan lines show the $M_{\ast}$--IRX relations from \protect\citealt{Qin2019} ($z=0.1$) and \protect\citealt{Whitaker2014} ($0.5\leq z\leq 2.5$), while the red lines represent the best-fitting $M_{\ast}$--IRX relations of the EAGLE SFGs in the same form as \protect Equations~\ref{eq:eq6}.}\label{fig:fig4}
\end{figure*}

\subsection{Correction to Galaxy Parameters}\label{sec:sec3.2}

The scaling relations from observations are used to examine physical parameters/quantities of the EAGLE SFGs, and quantify the discrepancies of the parameters/quantities at a given stellar mass. We adopt the $M_{\ast}$--$Z$ relation from \citet{Sanders2023}, the $M_{\ast}$--$R_{\rm e}$ relation from \citet{vanderWel2014}, the $M_{\ast}$--SFR relation from \citet{Popesso2023}, and the $M_{\ast}$--IRX relation from \citet{Whitaker2014} and \citet{Qin2019} for further analysis. 

\subsubsection{Gas-Phase Metallicity}\label{sec:sec3.2.1}

\citet{Sanders2023} determined the stellar mass--metallicity relation across cosmic time using observational data from the MOSDEF survey and the Sloan Digital Sky Survey. The $M_{\ast}$--$Z$ relation is parameterized using the formula 
\begin{equation}
\label{eq:eq2}
\begin{split}
& 12 + \log {\rm (O/H)} = Z_0 - \gamma/\beta \times \log (1 + {[M_{\ast}/{M_0}(z)]}^{-\beta}), \mathrm{with} \\
& \log ({M_0}(z)/{\rm {M_\odot}}) = m_0 + m_1 \times \log (1 + z),
\end{split}
\end{equation}
where $Z_0$ is the asymptotic metallicity at the high-mass end, $\gamma$ is the  power-law index at the low-mass end, $\beta$ controls the sharpness of the turnover that occurs at $M_0(z)$, and $m_0$ and $m_1$ are fitting parameters.
The best-fitting parameters given in Appendix B of \citet{Sanders2023} are adopted to describe the observed $M_{\ast}$--$Z$ relation.

Equation~\ref{eq:eq2} is used to fit the EAGLE SFGs at six redshifts, yielding the best-fitting $Z_0$, $\gamma$, $\beta$, $m_0$ and $m_1$. The corresponding relation is shown in Fig.~\ref{fig:fig3} via the red solid line. A constant offset in metallicity at a given stellar mass can be calculated between the observations (blue lines) and the EAGLE simulations (red lines) at each redshift. By adding the corresponding offset, the EAGLE SFGs are assigned the corrected gas-phase metallicity.

Fig.~\ref{fig:fig3} shows the comparison of the $M_{\ast}$--$Z$ relation between the original and corrected gas-phase metallicity. Statistically speaking, the EAGLE SFGs exhibit higher metallicity than the observations at all six redshifts. The discrepancy in gas-phase metallicity between the EAGLE simulations and the observations gradually decreases with increasing redshift. Note that we extrapolated the $M_{\ast}$--$Z$ relation from \citet{Sanders2023} to the low-mass regime ($10^{8.5-9.5}$ $\rm {M_\odot}$), and might potentially induce some bias \citep{Curti2020, Sanders2021}. We note that the systematic offset of gas-phase metallicity in observations can reach $\sim$0.7 dex at similar $z$ due to the effect of different samples, selection criteria, apertures and, especially, metallicity calibrators \citep{Genzel2015, DeRossi2017, Blanc2019}.

\subsubsection{Half-Light Radius}\label{sec:sec3.2.2}

Galaxy size (i.e. half-light radius) is another important parameter in determining IRX. We use the $M_{\ast}$--$R_{\rm e}$ distribution of late-type galaxies from \citet{vanderWel2014} to calibrate our EAGLE SFGs. \citet{vanderWel2014} presented the evolution of the $M_{\ast}$--$R_{\rm e}$ relation over a wide redshift range of $0 < z < 3$ for galaxies with $M_{\ast}>10^9$\,M$_\odot$ using data from the 3D-HST/CANDELS survey. The late-type galaxies have increasing $R_{\rm e}$ at increasing stellar mass at a given redshift, while at increasing redshift they appear to be substantially smaller than equally massive, present-day counterparts. The overall slope of the $M_{\ast}$--$R_{\rm e}$ relation shows no or little evolution. They parameterized the $R_{\rm e}$ as a function of stellar mass at six redshift bins following \citet{Shen2003}:
\begin{equation}
\label{eq:eq3}
R_{\rm e}/{\rm kpc} = A \cdot {\left(\frac{M_\ast}{5 \times 10^{10}{\rm M_\odot}}\right)}^\alpha,
\end{equation}
where $A$ and $\alpha$ are free parameters to fit data points. The best-fitting results were given in the table~1 of \citet{vanderWel2014}.

Because the redshift bins studies in \citet{vanderWel2014} is inconsistent with the six redshifts in our work, we use the spline interpolation method to derive the $M_{\ast}$--$R_{\rm e}$ relation at our given redshifts. Again, we extrapolate the $M_{\ast}$--$R_{\rm e}$ relation from \citet{vanderWel2014} down to $M_\ast=10^{8.5}$\,M$_\odot$. For the sake of acquiring the $R_{\rm e}$ offset between the EAGLE simulations and the observations, we fit the $M_{\ast}$--$R_{\rm e}$ distribution of the EAGLE SFGs using Equation~\ref{eq:eq3}. Similar to the 
offset correction for metallicity, we calculate the $R_{\rm e}$ offset for our EAGLE SFGs. The results are shown in Fig.~\ref{fig:figA2}. The $R_{\rm e}$ correction is strong function of galaxy stellar mass and redshift, because the EAGLE simulations produce low-mass SFGs larger and massive SFGs smaller compared with the observations, and this deviation gradually increases at increasing redshift.

Indeed, the deviation in size is so large that the slope of the best-fitting $M_{\ast}$--$R_{\rm e}$ relation even becomes negative at $z=2.5$ for the EAGLE SFGs, opposite to the positive slope of the observed $M_{\ast}$--$R_{\rm e}$ relation at all redshifts. We point out that the conversion of $R_{\rm e,mass}$ from the EAGLE simulations to $R_{\rm e}$ may cause some uncertainties, and the increasing scatter in $R_{\rm e}$ with redshift and stellar mass might also contribute to the $R_{\rm e}$ deviation.

\begin{figure*}
	\centering
	\includegraphics[width=0.9\textwidth]{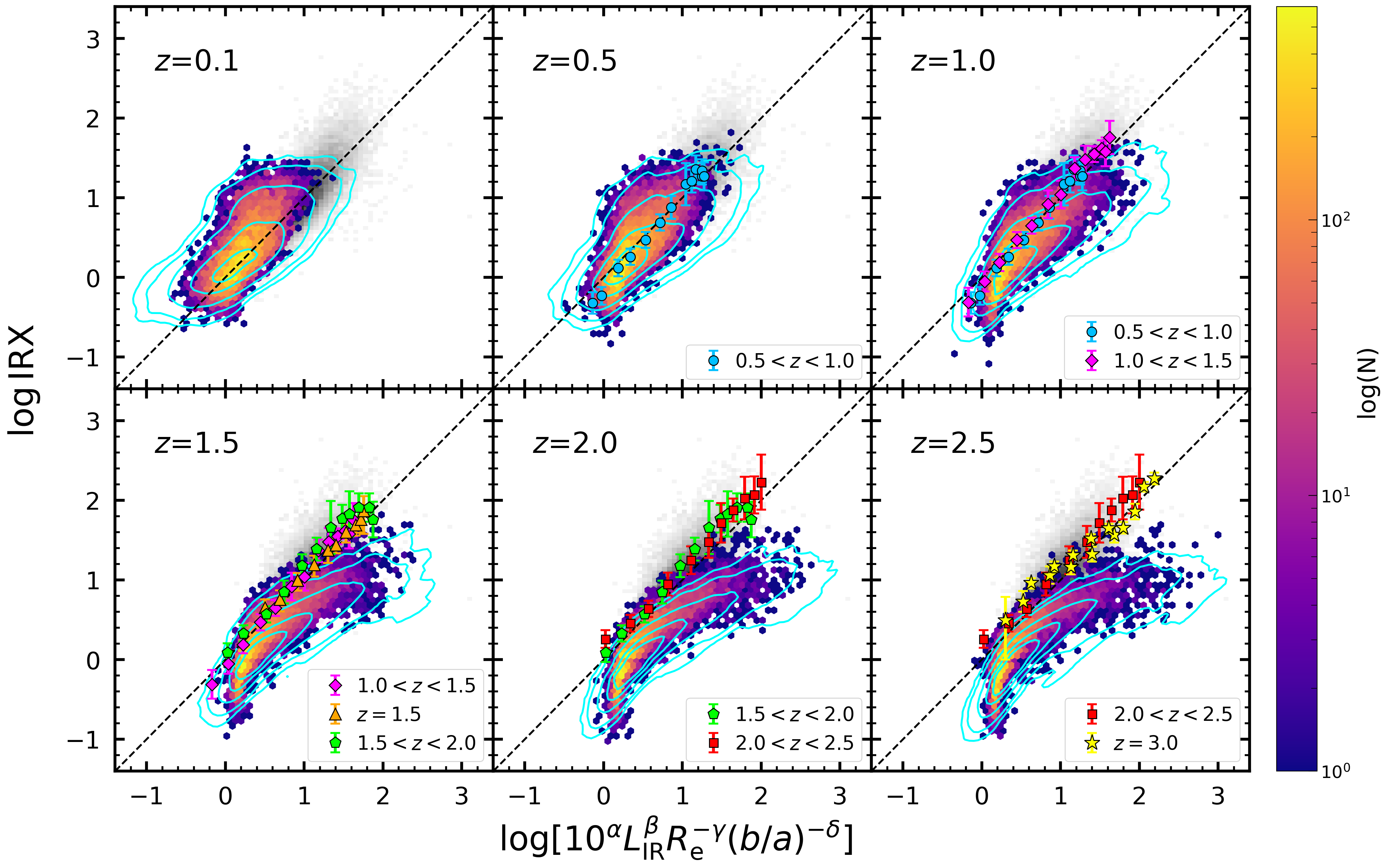}
    \caption{The predicted universal IRX relations of the EAGLE SFGs with corrected metallicity  at six redshifts. The gray-scale map demonstrates the distribution of local SFGs and the data points with error bars refer to the average SFGs of different stellar masses at given redshifts as described in Fig.~\ref{fig:fig2}. the colour of the bins represents the number of SFGs within that bin. The cyan contours show the distributions of the EAGLE SFGs with original parameters, and the outermost contour is the same as in Fig.~\ref{fig:fig2}. }\label{fig:fig5}
\end{figure*}

\subsubsection{IR Luminosity and IR Excess}\label{sec:sec3.2.3}

The third important parameter to be corrected for our EAGLE SFGs is $L_{\rm IR}$. Consequently IRX needs to be corrected as well. In practice, we firstly calibrate SFR, and secondly IRX. The corrections for SFR and IRX determine the correction to $L_{\rm IR}$.

Compiling a comprehensive census of literature studies, \citet{Popesso2023} investigated the evolution of the $M_{\ast}$--SFR relation of SFGs with $M_\ast=10^{8.5-11.5}$\,M$_\odot$ in the redshift range $0<z<6$. They adopted the same form as \citet{Lee2015} to parameterize the $M_{\ast}$--SFR relation:
\begin{equation}
\label{eq:eq4}
{\rm SFR} = \frac{\rm SFR_{\max}}{1 + {(M_{\ast}/M_0)}^{-\gamma}},
\end{equation}
where SFR$_{\max}$ is given in unit of M$_\odot$\,yr$^{-1}$ and refers to the maximal SFR that the relation reaches at the high-mass end, $M_0$ is the turnover stellar mass and $\gamma$ is the power-law slope at the low-mass end. Here SFR$_{\max}$ and $M_0$ depend on time, following:
\begin{equation}
\label{eq:eq5}
\begin{split}
& \log {\rm SFR}_{\max} = a_0 + a_1t, \\
& \log M_0 = a_2 + a_3t.
\end{split}
\end{equation}
The $M_{\ast}-$SFR relation bends at the high-mass end with the turnover mass $M_0$ evolving with time which is possibly a result of star formation quenching.

The best-fitting $M_{\ast}$--SFR relation for the EAGLE SFGs is derived using Equation~\ref{eq:eq4} and Equation~\ref{eq:eq5}, and shown in Fig.~\ref{fig:figA3}. We compare the best-fitting relation of the EAGLE SFGs with the $M_{\ast}$--SFR relation from \citet{Popesso2023} to calculate the offset in SFR between the EAGLE simulations and the observations at given stellar masses and redshifts. We then correct the SFR offset for the EAGLE SFGs, and obtain the corrected SFR, as shown in Fig.~\ref{fig:figA3}. It is clear that the EAGLE SFGs exhibit systematically lower SFR compared to the observed SFGs across all stellar masses and redshifts. The offset between \citet{Popesso2023} and this work gradually increases from a minimum of $\sim$0\,dex to a maximum of $\sim$0.6\,dex at increasing redshift. Corrected for the SFR offset, the EAGLE SFGs follow the observed $M_{\ast}$--SFR relation, which demonstrates the difference of the normalization between the observed and simulated relation. We note that there is strong evidence that the $M_{\ast}$--SFR relation is dominated by systematics. Studies that calculated the SFR from a combination of UV+IR luminosity have overestimated normalization \citep{Katsianis2020,Leja2022}, and thus the above tension between the EAGLE simulations and the observations can be attributed to limitations of the determination of galaxy properties.

For the next step, we calibrate the IRX for our EAGLE SFGs using the $M_{\ast}$--IRX relation from \citet{Whitaker2014} and \citet{Qin2019}. \citet{Whitaker2014} selected 39,106 SFGs with 0.5 < $z$ < 2.5 from the 3D-HST/CANDELS survey, and performed a stacking analysis of the Spitzer/MIPS imaging data. They found that the SFGs with $\log(M_{\ast}/$M$_\odot)<10.5$ showed no evolution in IRX from $z=0.5$ to $z=2.5$ for SFGs, while SFGs with $\log (M_\ast$/M$_\odot)>10.5$ appeared to have a higher IRX at increasing redshift. We adopt the average $L_{\rm IR}$ and $L_{\rm UV}$ measurement for subsamples in given redshift and stellar mass bins from \citet{Whitaker2014}, and fit the $M_{\ast}$--IRX distribution at given redshift with a second-order polynomial function as
\begin{equation}
\label{eq:eq6}
\log {\rm IRX} = a + b\log \left(\frac{M_{\ast}}{\rm {M_\odot}}\right) + c\log {\left(\frac{M_{\ast}}{\rm {M_\odot}}\right)}^2.
\end{equation}
And we apply the spline interpolation method to obtain the $M_{\ast}$--IRX relation for our five redshifts (except $z=0.1$). In addition, for $z=0.1$, we fit the data from \citet{Qin2019} to obtain the observed $M_{\ast}$--IRX relation. Using Equation~\ref{eq:eq6}, We derive the best-fitting $M_{\ast}$--IRX relation of our EAGLE SFGs at six redshifts. Doing so we calculate the IRX offset between the EAGLE simulations and the observations, and correct our EAGLE SFGs for the offset and present the correction results in Fig.~\ref{fig:fig4}.

It is clearly demonstrated that the EAGLE SFGs show statistically lower IRX than the observed SFGs at $z=0.1$. However, at $z\geq 0.5$, the EAGLE SFGs show a trend of higher IRX at the low-mass end and lower IRX at the high-mass end compared with the observations, causing an intersection in the best-fitting relation between the EAGLE simulations and the observations, and the stellar mass corresponding to the intersection increases with increasing redshift.

Using ${\rm SFR}=1.09\times10^{10}\,(L_{\rm IR}+2.2\,L_{\rm UV}$) \citep{Bell2005}, we are able to obtain the expected $L_{\rm IR}$ and $L_{\rm UV}$ from the corrected SFR and corrected  IRX ($=L_{\rm IR}/L_{\rm UV}$). Then we employ a second-order polynomial to fit the $M_{\ast}$--$L_{\rm IR}$ relation for the EAGLE SFGs with original and expected (or corrected) $L_{\rm IR}$ at six redshifts, and the offset is shown in Fig.~\ref{fig:figA4}. It is obvious that the overall tendency of the discrepancy between the original and corrected $M_{\ast}$--$L_{\rm IR}$ relation mirrors that of the $M_{\ast}-$IRX relation in Fig.~\ref{fig:fig4}.

\begin{figure*}
	\centering
	\includegraphics[width=0.95\textwidth]{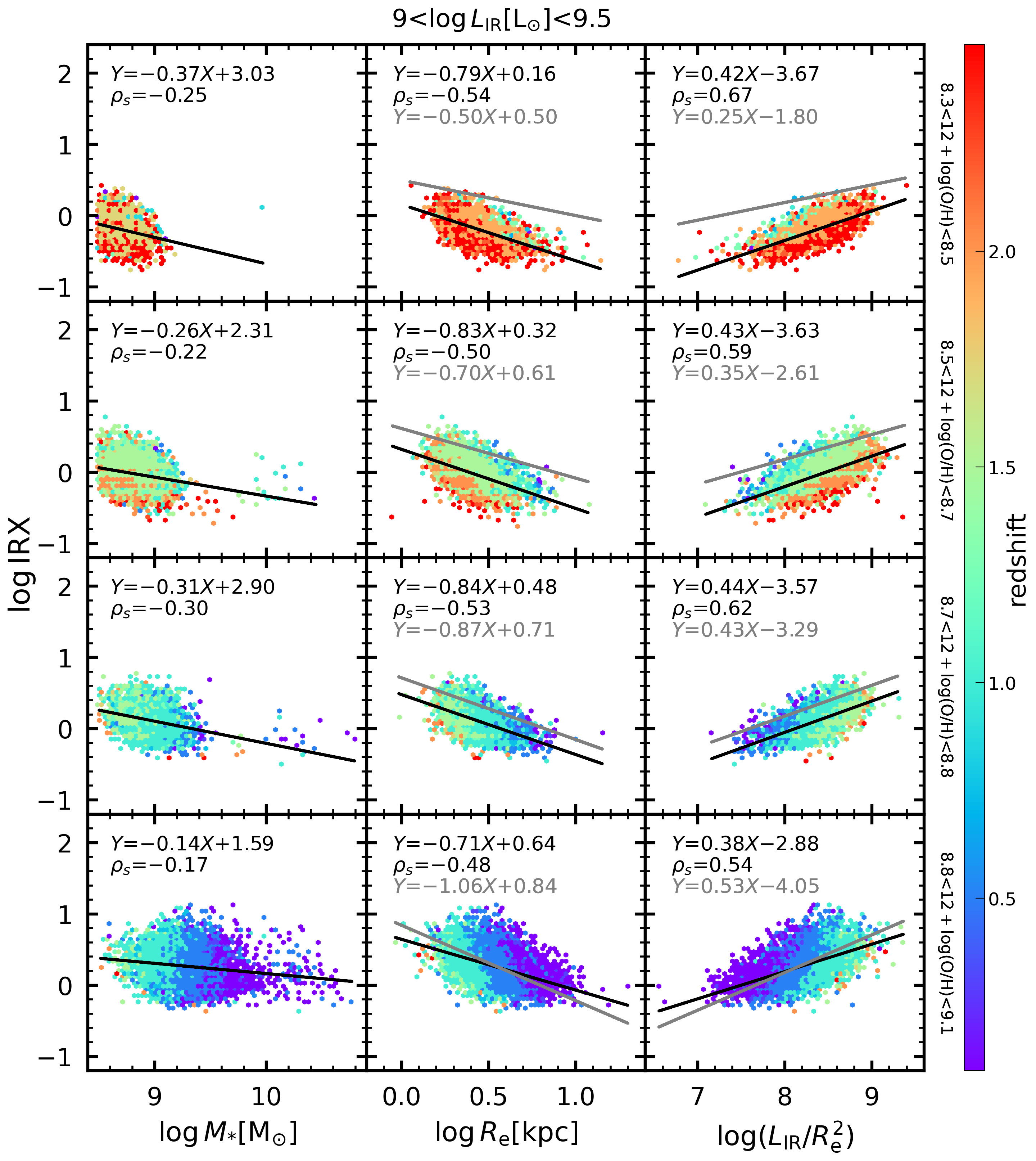}
    \caption{Relationships between IRX and $M_{\ast}$, $R_{\rm e}$ and $L_{\rm IR}/R_{\rm e}^2$ (from left to right) for our EAGLE SFGs with $9<\log (L_{\rm IR}/{\rm L_\odot}) <9.5$ split into four metallicity bins over $8.3<12 + \log {\rm (O/H)} <9.1$ (from top to bottom). The colour of the bins is coded by the median redshift of the EAGLE SFGs within that bin. The black and grey solid lines denote the best-fitting power-law relations in this work and from \protect\citet{Qin2019}, and the relevant best-fitting functions are shown at the top of each panel. The coefficient of Spearman's rank $\rho_{\rm s}$ is presented to show the strength of the correlation.}\label{fig:fig6}
\end{figure*}

\subsection{The Predicted Universal IRX Relation with Corrected Parameters}\label{sec:sec3.3}

The predicted universal IRX relation of the EAGLE SFGs based on the original galaxy parameters shows increasing deviation from the observed relation at $z\geq 1.0$, as shown in Fig.~\ref{fig:fig2}. Naturally, we intend to figure out the causes for the discrepancy between the EAGLE simulations and the observations. For this purpose, we use the scaling relations ($M_{\ast}$--$Z$, $M_{\ast}$--$R_{\rm e}$, $M_{\ast}$--SFR and $M_{\ast}$--IRX relation) of observed SFGs to re-calibrate the simulated galaxy parameters in order to investigate which parameters in the EAGLE SFGs significantly contribute to the discrepancy. In each of the following sections, the predicted universal IRX relation is re-examined with changes to only one parameter among metallicity, $R_{\rm e}$ and $L_{\rm IR}$ (hence IRX).

\subsubsection{The Gas-Phase Metallicity-related Correction}\label{sec:sec3.3.1}

We firstly study the influence of $Z$ on the predicted universal IRX relation. Fig.~\ref{fig:fig5} shows the distribution of the EAGLE SFGs with only correction of parameter $Z$ in the universal IRX relation. We notice that the EAGLE SFGs with $\log {\rm IRX} > 0$ are mostly affected by the metallicity correction, resulting in a better agreement with the observed universal IRX relation. This improvement indicates that the gas-phase metallicity is a key quantity in reproducing the observed universal IRX relation. However, for the low-IRX galaxies ($\log {\rm IRX} < 0$), the deviation becomes larger, especially at high redshift. We argue that this increasing deviation might be biased by the correction to metallicity, because the extrapolation of the observed $M_{\ast}$--$Z$ relation further into the low-mass and low-metallicity regime is of relatively high uncertainty. This uncertainty could be magnified in the extrapolation to the very low-IRX end, even leading $\beta$ in Equation~\ref{eq:eq1} to be negative, and consequently stronger deviation in the universal IRX relation.

\subsubsection{The $R_{\rm e}$-related Correction}\label{sec:sec3.3.2}

After considering the gas-phase metallicity, galaxy size $R_{\rm e}$ is another parameter to be addressed. Comparison of the predicted universal IRX relation with the original parameters versus the parameters with only correction of $R_{\rm e}$ is shown in Fig.~\ref{fig:figA5}. The correction to $R_{\rm e}$ is dramatic as shown in Fig.~\ref{fig:figA2}. However, the predicted universal IRX relation at six redshifts all change little when replacing the original Re with the one with correction. This result suggests that the Re is not a dominant parameter responsible for the discrepancy of the universal IRX relation between the EAGLE simulations and the observations.

\begin{figure*}
	\centering
	\includegraphics[width=0.98\textwidth]{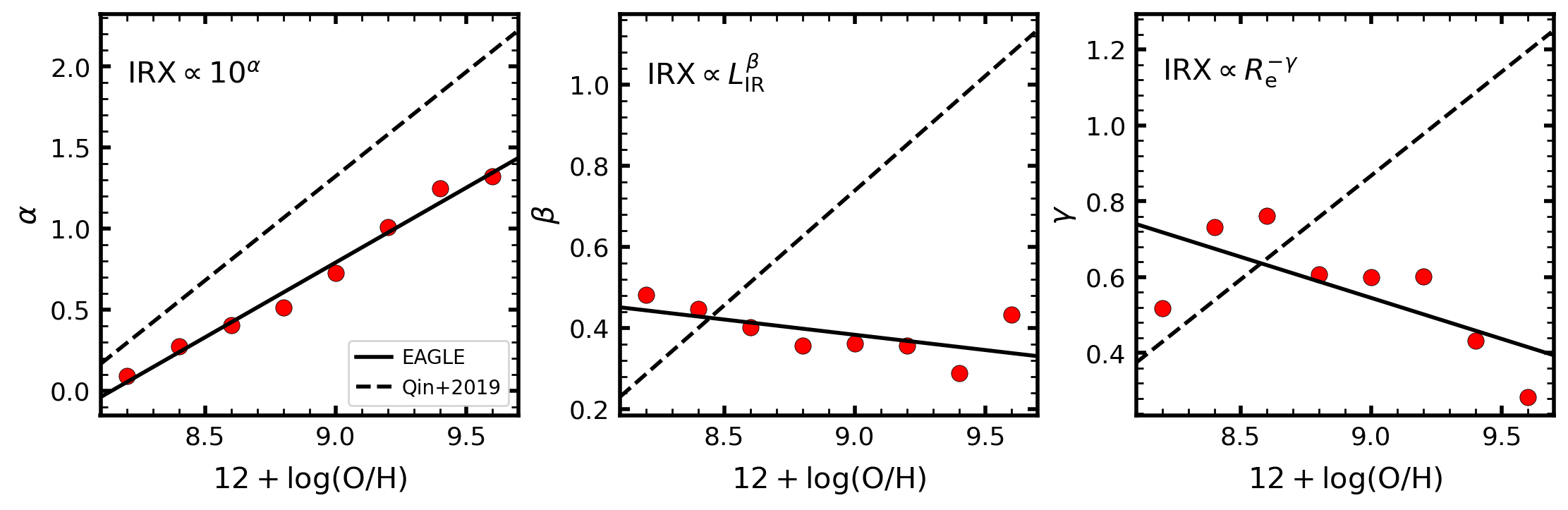}
    \caption{Power-law indices $\alpha$, $\beta$ and $\gamma$ as a function of gas-phase metallicity for the sample SFGs with the original galaxy parameters from the EAGLE simulations. The red solid circles represent the best-fitting values. The solid and dashed lines refer to the best-fitting relation in our work and from \protect\citet{Qin2019}, respectively.}\label{fig:fig7}
\end{figure*}

\subsubsection{The $L_{\rm IR}$-related and IRX-related Correction}\label{sec:sec3.3.3}

The third parameter we examine for the predicted universal IRX relation is the IR luminosity. Because IRX contains $L_{\rm IR}$, variation in $L_{\rm IR}$ will have obviously effect on IRX. We thus determine their impact on the predicted universal IRX relation together, and the result with corrections of $L_{\rm IR}$ and IRX are illustrated in Fig.~\ref{fig:figA6}. As can be seen, the positions of the EAGLE SFGs remain nearly unchanged after implementing the corrections to $L_{\rm IR}$ and IRX, and those with $\log$ IRX < -0.5 show even a slightly larger discrepancy. We point out that the corrections of $L_{\rm IR}$ and IRX for the EAGLE SFGs are based on the scaling relations from observational studies, and the extrapolation to the low-mass regime might suffer from large uncertainties, similar to the situation for the gas-phase metallicity discussed in Section~\ref{sec:sec3.3.1}. We thus argue that the impact of $L_{\rm IR}$ and IRX on the predicted universal IRX relation is likely marginal.

\subsection{Relationships between Original IRX and Galaxy Parameters}\label{sec:sec3.4}

In order to understand the discrepancy of the universal IRX relation between the EAGLE simulations and the observations, we investigate the correlations between the original IRX and original galaxy parameters from the EAGLE simulations and compared these correlations with that of the observations. For the reason that $M_{\ast}$, SFR, $Z$ and $L_{\rm IR}$ are tightly correlated with each other, we examine the relationships between the IRX and galaxy parameters at fixed $L_{\rm IR}$ and $Z$. This approach is similar to that adopted in \citet{Qin2019}.

We analyze a subsample of the EAGLE SFGs with $9 < \log (L_{\rm IR}/{\rm L_\odot}) < 9.5$, and divided them into four metallicity bins over $8.3 < 12 + \log {\rm (O/H)} < 9.1$. Here the $L_{\rm IR}$ range is selected to maximize the number of sample galaxies for good statistics and meanwhile to minimize the property changes with $L_{\rm IR}$; then the subsample is divided into four metallicity bins over $8.3 < 12 + \log {\rm (O/H)} < 9.1$ to allow each bin contains a similar number of SFGs. Fig.~\ref{fig:fig6} shows the correlations of IRX with $M_{\ast}$, $R_{\rm e}$ and $L_{\rm IR}/R_{\rm e}^2$ as a function of gas-phase metallicity. Our analysis reveals that for all four metallicity bins, IRX decreases with $R_{\rm e}$ and increases with $L_{\rm IR}/R_{\rm e}^2$. A rather weak anti-correlation between $M_{\ast}$ and IRX is presented, while in observations IRX is independent of $M_{\ast}$ \citep{Qin2019}. We argue that the weak dependence of IRX on $M_{\ast}$ is largely induced by the systematic discrepancies of galaxy parameters for the EAGLE SFGs. Additionally, for a given $M_{\ast}$, $R_{\rm e}$ or $L_{\rm IR}/R_{\rm e}^2$, the average IRX increases with increasing gas-phase metallicity. These findings are consistent with the results given in \citet{Qin2019}. However, we see that the slope of the best-fitting relation for IRX with $R_{\rm e}$ or $L_{\rm IR}/R_{\rm e}^2$ does not become steeper at increasing gas-phase metallicity in the EAGLE simulations, differing from the tendency reported by \citet{Qin2019}.

We also investigate the correlations between IRX and $Z$, $R_{\rm e}$, $L_{\rm IR}$ and $L_{\rm IR}/R_{\rm e}^2$ while fixing $L_{\rm IR}$ and $M_{\ast}$ or $M_{\ast}$ and $Z$, as shown in Figures~\ref{fig:figA7} and~\ref{fig:figA8}. We point out that the discrepancy in $L_{\rm IR}$ (also IRX) contributes more than that of $R_{\rm e}$ in responsible for the bigger discrepancy of the universal IRX relation at $z\geq 1.0$ between the EAGLE simulations and the observations.  We conclude that the discrepancy between the EAGLE simulations and the observations may be due to the fact that the EAGLE simulations are not able to reproduce the slopes of the relationships between IRX and $R_{\rm e}$ and $L_{\rm IR}/R_{\rm e}^2$, which increase with gas-phase metallicity.

\subsection{The Dependence of Power-law Indices of the Predicted Universal IRX Relation on Gas-phase Metallicity}\label{sec:sec3.5}

We further investigate the slope of the power-law indices of the predicted universal IRX relation as a function of gas-phase metallicity for the EAGLE SFGs with the original galaxy parameters. The EAGLE SFGs in our sample are divided into eight continuous metallicity bins over $8.1 < 12 + \log ({\rm O/H}) < 9.7$, and fitted with the universal IRX relation described by Equation~\ref{eq:eq1} in each gas-phase metallicity bin. The relations between gas-phase metallicity and power-law indices ($\alpha$, $\beta$ and $\gamma$) are examined. We do not include $\delta$ because we set inclination angle to be constant ($b/a=0.6$). Fig.~\ref{fig:fig7} shows the best-fitting power-law indices as a function of gas-phase metallicity. We demonstrate that only $\alpha$ increases with gas-phase metallicity, with a slope consistent with that from the observations, whereas $\beta$ and $\gamma$ decrease with gas-phase metallicity, dramatically differing from those of the observations. These results reveal some limitations of the EAGLE simulations which are not able to reproduce the dependence of the IRX on $R_{\rm e}$ and $L_{\rm IR}$. Further efforts are needed to improve the simulations in reproducing the gas-phase metallicity and dust attenuation for the SFGs across cosmic time.

\section{Discussion and Prospects for State-of-the-art Cosmological Simulations}\label{sec:sec4}

Provided that the EAGLE simulations successfully captured qualitatively numerous scaling relations and the evolution of galaxy properties like the SFR and stellar mass, in our study we decide to perform comparisons of the reference model Ref-L0100N1504 with the observations in the universal IRX relation. Our study is aiming to unravel the performance of state-of-the-art simulations with respect additional observational data and uncover potential areas of discrepancy, such as gas-phase metallicity, size, infrared luminosity ($L_{\rm IR}$) and IRX.

The agreement between the EAGLE simulations and the observations in the universal IRX relation up to $z=0.5$ compliments the simulation's capacity to replicate certain important aspects of galaxy formation and evolution, especially at low redshift. However, at $z\geq 1.0$ the deviations between the predicted and observed universal IRX relations become pronounced, particularly for galaxies at both the low-IRX and high-IRX end.

Previous studies like \citet{Furlong2017} investigated the evolution of the half-mass radius for galaxies from $z=2$ to $z=0$ in the EAGLE simulations, and made a comparison with the half-light radius $R_{\rm e}$ inferred from the observations of \citet{Shen2003} and \citet{vanderWel2014}. The agreement for galaxies with $M_\ast<10^{11}$\,M$_\odot$ was within 0.1\,dex at $z<1$ and 0.2--0.3\,dex at $1\leq z\leq 2$. However, in our work we caution that the comparison of the half-mass radius with the half-light radius may introduce a systematic bias because of the presence of a colour gradient (or stellar-age gradient) in galaxies. It seems that the sizes of the EAGLE galaxies are biased in the sense that low-mass galaxies are too extended and high-mass galaxies are too compact, even leading to a negative slope for the $M_{\ast}$--$R_{\rm e}$ relation at $z=2.5$, as shown in Fig.~\ref{fig:figA2}.

\citet{DeRossi2017} also discussed the correlations between the galaxy stellar mass and gas-phase metallicity in the EAGLE simulations. They argued that the gas-phase metallicity in massive galaxies ($M_{\ast} \gtrsim 10^{10}$\,M$_\odot$) was mostly regulated by the AGN feedback whereas the slope of the simulated $M_{\ast}$--$Z$ relation was mostly determined by stellar feedback in low-mass galaxies ($M_{\ast} \lesssim 10^{10}$\,M$_\odot$). Increasing the AGN feedback heating temperature yields decrease of the gas-phase metallicity at the high-mass end, by heating the gas remaining, thereby decreasing SFR and by ejecting metal-enriched gas from galaxies. Weaker stellar feedback tends to produce higher gas-phase metallicity in low-mass galaxies. However, still the discrepancies between different observational results can reach $\sim$0.7\,dex at similar redshift due to the effect of different samples, selection criteria, apertures and, especially, metallicity indicators affecting our comparison and qualitative results \citep{DeRossi2017}.

\citet{Furlong2015} studied the evolution of galaxy stellar mass and SFR in the EAGLE simulations up to $z=2$, and found that sSFR of SFGs was typically 0.2--0.5\,dex lower than that from the observations at all redshifts and across all stellar masses. The same results were obtained by \citet{Schaye2015} at low redshifts. Besides that the $M_{\ast}$--SFR relation is subject to systematics and biases \citep{Katsianis2020}, it is suspected that insufficient bursts of star formation in the simulations leave sSFR lower for simulated galaxies, something that should be addressed. We note that star formation in simulations usually has a sophisticated metallicity-dependent density threshold, and only gas particles above this threshold are likely to form stars, determined by their pressure.

A significant result of our study is the application of the parameter corrections to the EAGLE galaxy parameters, including the gas-phase metallicity, galaxy size ($R_{\rm e}$), $L_{\rm IR}$ and IRX, to better align with the observational trends. Notably, the corrections related to gas-phase metallicity emerge as the primary driver of deviations in the predicted universal IRX relation. The gas-phase metallicity decides the power indices of four terms in the universal IRX relation given in Equation~\ref{eq:eq1}, and thus the correction to metallicity induces changes to all of the four terms. In contrast, the correction to either $R_{\rm e}$ or $L_{\rm IR}$ only leads to change related a single term. Our results have shown that the predicted and observed IRX relations globally agree with each other at $z\leq 0.5$ and deviation between the two becomes increasingly larger at increasing redshift. The deviation is jointly contributed by the systematic discrepancies of multiple galaxy parameters, particularly for the EAGLE SFGs at the low-IRX and high-IRX ends. There are some severe disparities in the power indices of the universal IRX relation in the EAGLE simulations and the observations. These indices hold immense promise as a valuable diagnostic tools for refining hydrodynamical cosmological simulations like EAGLE, offering the potential to resolve simultaneously issues associated with star formation rates, chemical enrichment processes, and the evolving structural characteristics of galaxies.

Looking ahead, we make suggestions for the next-generation cosmological simulations. A possibility of the discrepancies highlighted in our study is that the geometry of the dust and stars is not considered/resolved in the EAGLE simulations, a limitation that is common among current cosmological simulations performing over large volumes,  although the distribution of stars and dust is to some extent resolved through the particles or resolution elements. This can be a severe limitation at low redshifts since with increasing gas-phase metallicity the molecular clouds become  larger, and consequently the dust attenuation of galaxies. In contrast, SED fitting usually imposes a star-dust geometry through the choice of attenuation curves. The IRX will depend on this choice as well if the SED does not cover the infrared. Simulations with higher resolution (including a range of high resolution zoom-in simulations) may address the issue.  Second, we note that models may need to generate galaxies with elevated star formation rates and low metallicities, particularly at high redshifts. More galaxy inflows and more outflows at high redshifts could lead to a decrease of the gas-phase metallicity and hence higher star formation rates. However, these changes are not trivial to implement due to the fact that there is a complex interplay between these processes \citep{Agertz2015, Tescari2014, Blanc2019, Jecmen2023}. In addition, it is challenging to reproduce some key observables simultaneously. For example, by decreasing feedback at higher redshifts, higher star formation rates would be achieved and indeed a better agreement between IR derived star formation rate functions and EAGLE could exist \citep{Katsianis2017b}. However, this change if not addressed properly will affect the comparison with the stellar mass function at $z=0-8$, which is well described by the original model \citep{Furlong2015}. Resolution effects are demonstrated to impact the results of cosmological simulations. This makes particularly challenging the implementation of sub-grid schemes and uncoupling the physics from the numerical limitations \citep{Zhao2020}. Last but not least, dust physics and dust grain evolution have not been followed in detail in the EAGLE simulations. We note that the integration of realistic dust physics and an improved understanding of dust attenuation within galaxies across a diverse array of parameters will likely be pivotal in enhancing the robustness of our simulations.

\section{SUMMARY}\label{sec:sec5}

In this study, we use the reference model Ref-L0100N1504 from the EAGLE simulations to investigate the predicted universal IRX relation in comparison with the observed one. We quantify the discrepancies of galaxy parameters between the EAGLE simulations and  the observations using the observed scaling relations ($M_{\ast}$--$Z$, $M_{\ast}$--$R_{\rm e}$, $M_{\ast}$--SFR and $M_{\ast}$--IRX relation), and examine which parameters cause the deviation between the predicted and the observed universal IRX relation. Moreover, we also address the dependence of IRX on the gas-phase metallicity and other quantities. We arrive at the following conclusions:

\begin{enumerate}[i)]

	\item The predicted universal IRX relation by the EAGLE simulations follows the observations at $z\leq 0.5$, but exhibits increasing deviation at increasing redshift, particularly for SFGs at the low-IRX and high-IRX end.

	\item We modify the predicted universal IRX relation from the EAGLE simulations by considering corrections to the individual galaxy parameter ($Z$, $R_{\rm e}$, or $L_{\rm IR}$ and IRX) related to it. We illustrates that the gas-phase metallicity holds the dominant role on the universal IRX relation, whereas the $R_{\rm e}$, $L_{\rm IR}$ and IRX only slightly impact the predicted universal IRX relation. This is consistent with the understanding based on the observational analysis given in \citet{Qin2019}. 

	\item Overall, besides its successes, the EAGLE simulations are not able to describe the dependence of IRX and other galaxy parameters on the gas-phase metallicity. This results in strong deviations for the power indices in the predicted universal IRX relation. We stress that these power indices can be used as effective probes for hydrodynamical cosmological simulations like EAGLE to reconcile issues related to star formation, chemical enrichment and structural evolution of galaxies in a single/consistent framework. 

\end{enumerate}

The EAGLE cosmological hydrodynamical simulations successfully reproduce many aspects of galaxies and structures in the Universe. The universal IRX relation is a powerful tool to test the simulations on modeling metal enrichment of the ISM in conjunction with star formation and structure growth of galaxies. Our investigations suggest that next generation sophisticated simulations should generate galaxies with higher star formation rate and lower metallicity at high redshift. This can be addressed by adding more realistic dust physics and reproducing dust attenuation for galaxies over wide parameter ranges.


\section*{Acknowledgements}

We thank the anonymous referee for her/his valuable comments and suggestions that improved this manuscript. This work is supported by the National Key Research and Development Program of China (2023YFA1608100), the National Natural Science Foundation of China (12073078, 12233005, 12173088 and 12033004), and the science research grants from the China Manned Space Project with NO. CMS-CSST-2021-A02, CMS-CSST-2021-A04 and CMS-CSST-2021-A07. AK thanks the 100 talents Program offered by the Sun Yat-sen University. We acknowledge the Virgo Consortium for making their simulation data available. The EAGLE simulations were performed using the DiRAC-2 facility at Durham, managed by the ICC, and the PRACE facility Curie based in France at TGCC, CEA, Bruy\'eresle-Ch\^atel.

\section*{Data Availability}

The data underlying this article are publicly available at http://www.eaglesim.org/database.php.




\bibliographystyle{mnras} 

\bibliography{EAGLE} 




\appendix

\section{Ancillary  Figures}

\begin{figure*}
    \centering
	\includegraphics[width=0.95\textwidth]{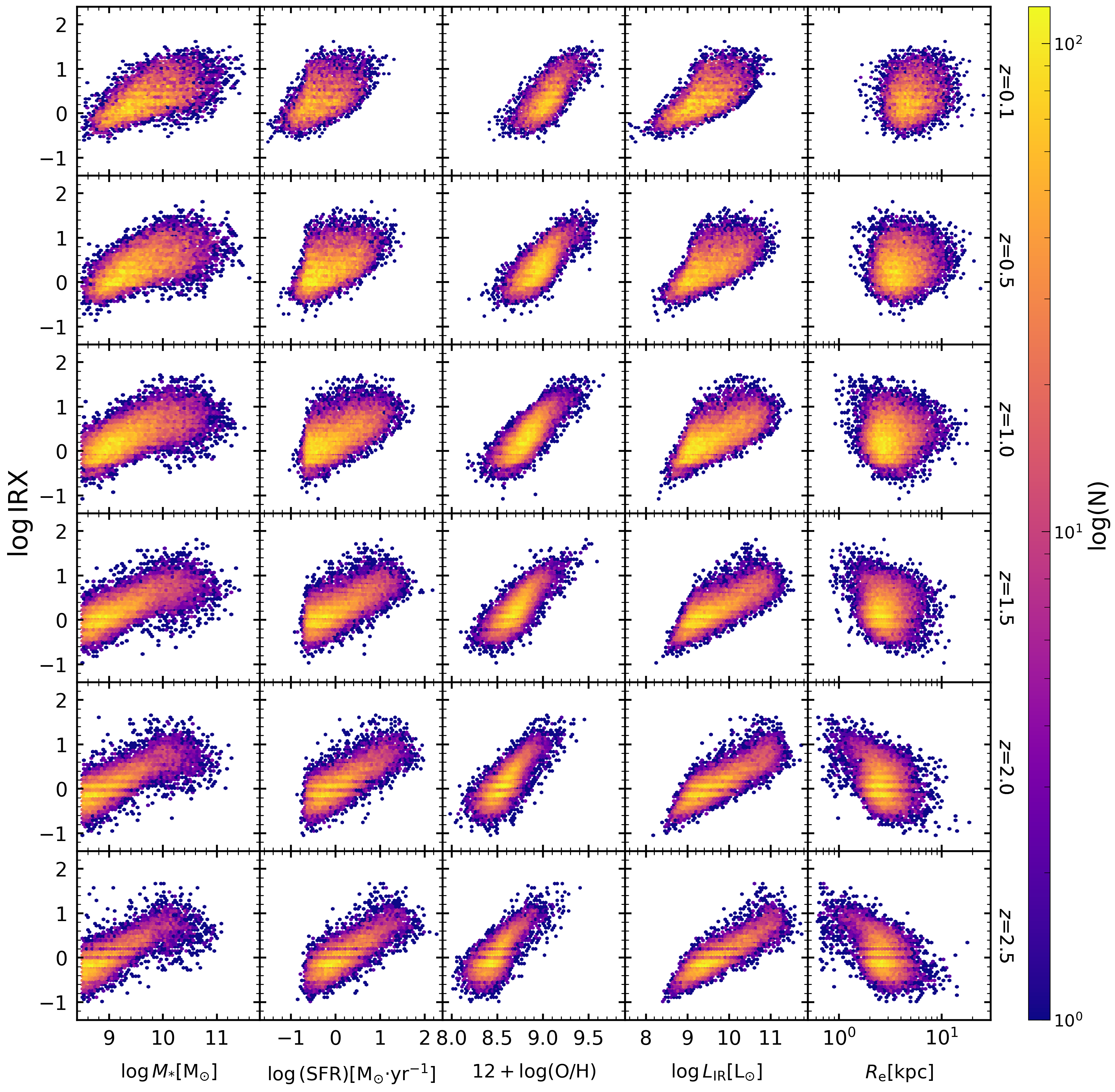}
    \caption{Relationships between $M_{\ast}$, SFR, $Z$, $L_{\rm IR}$, $R_{\rm e}$ and IRX for our sample EAGLE SFGs at six redshifts from $z=0.1$ to $z=2.5$. The colour of the hexagonal bins represents the number of data points within that bin.}\label{fig:figA1}
\end{figure*}

\begin{figure*}
	\centering
	\includegraphics[width=0.8\textwidth]{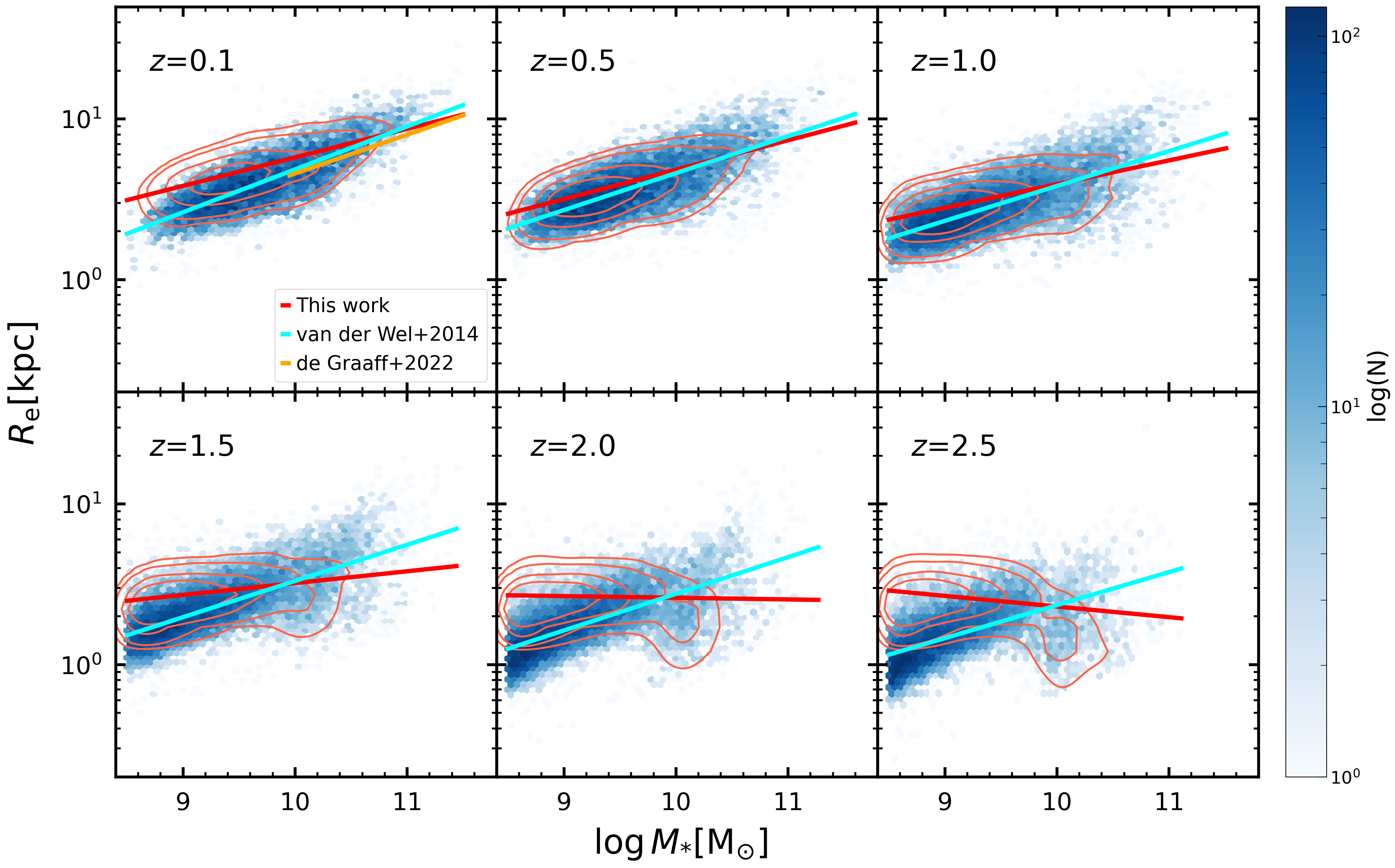}
    \caption{The stellar mass versus half-light radius relations at six redshifts for the EAGLE SFGs with original (red contours) and corrected (colour-coded hexagonal bins) $R_{\rm e}$. The colour of the hexagonal bins represents the number of the EAGLE SFGs with corrected $R_{\rm e}$ within that bin. The red lines represent the best-fitting relations given in the form of Equations~\ref{eq:eq3} to the data points of red contours, while the cyan lines refer to the $M_{\ast}$--$R_{\rm e}$ relations from \protect\citet{vanderWel2014} extrapolated to the low-mass end. The orange line reprents the $M_{\ast}$--$R_{\rm e}$ relation from \protect\citet{deGraaff2022}. Apparently, the EAGLE SFGs are overlarge at the low-mass end and too small at the high-mass end, leading the $M_{\ast}$--$R_{\rm e}$ relations to deviate mostly in slope from the observed relations at higher redshift.}\label{fig:figA2}
\end{figure*}

\begin{figure*}
	\centering
	\includegraphics[width=0.8\textwidth]{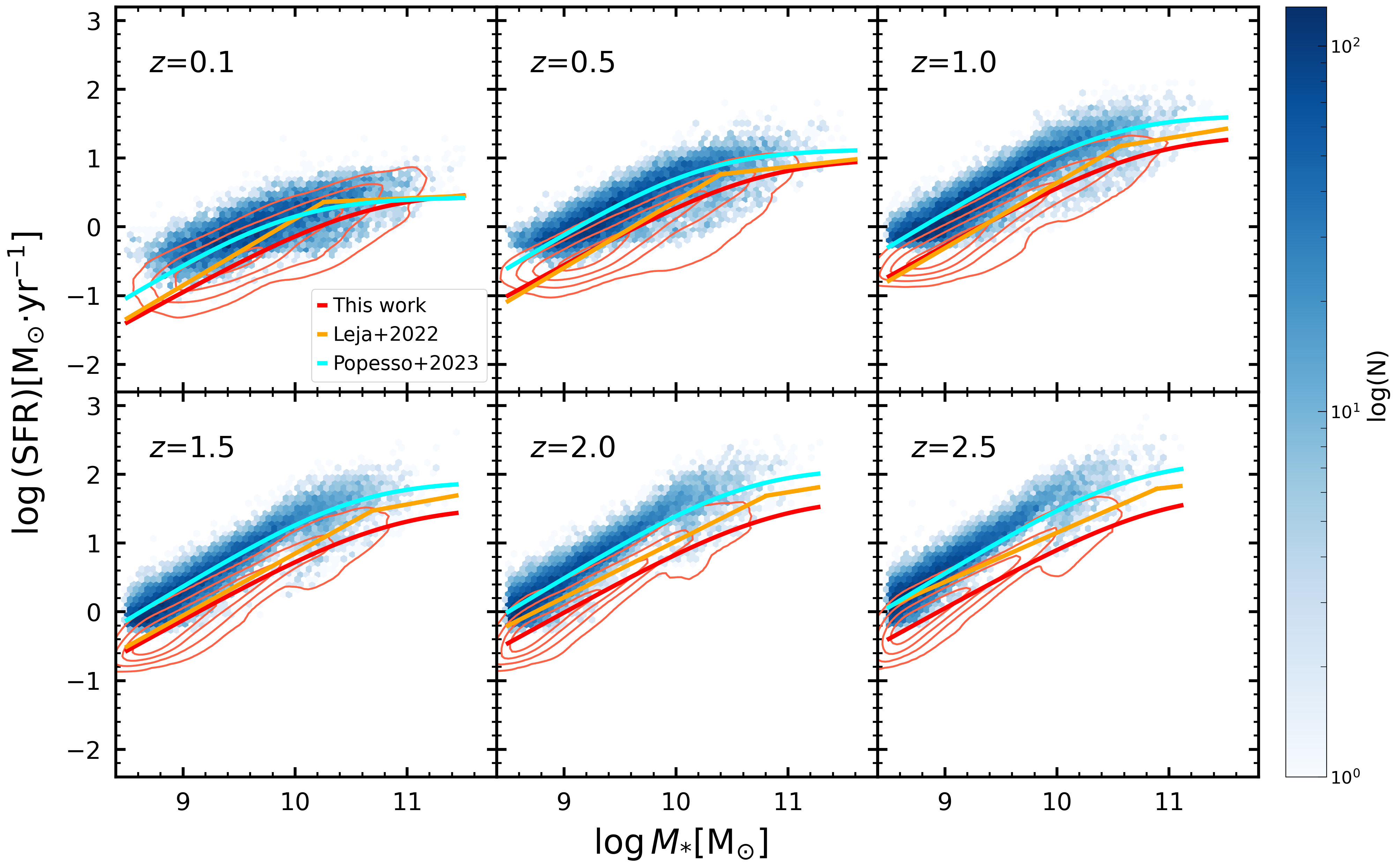}
    \caption{The stellar mass versus SFR relations at six redshifts for the EAGLE SFGs with original (red contours) and corrected (coloured hexagonal bins) SFR. Each hexagonal bin is colour coded by the number of the EAGLE SFGs within that bin. The cyan lines show the $M_{\ast}$--SFR relations from \protect\citet{Popesso2023}, whereas the red lines represent the best-fitting $M_{\ast}$--SFR relations for the EAGLE SFGs in the same form as the cyan lines. The orange lines represent the $M_{\ast}$--SFR relations from \protect\citet{Leja2022}.}\label{fig:figA3}
\end{figure*}

\begin{figure*}
	\centering
	\includegraphics[width=0.8\textwidth]{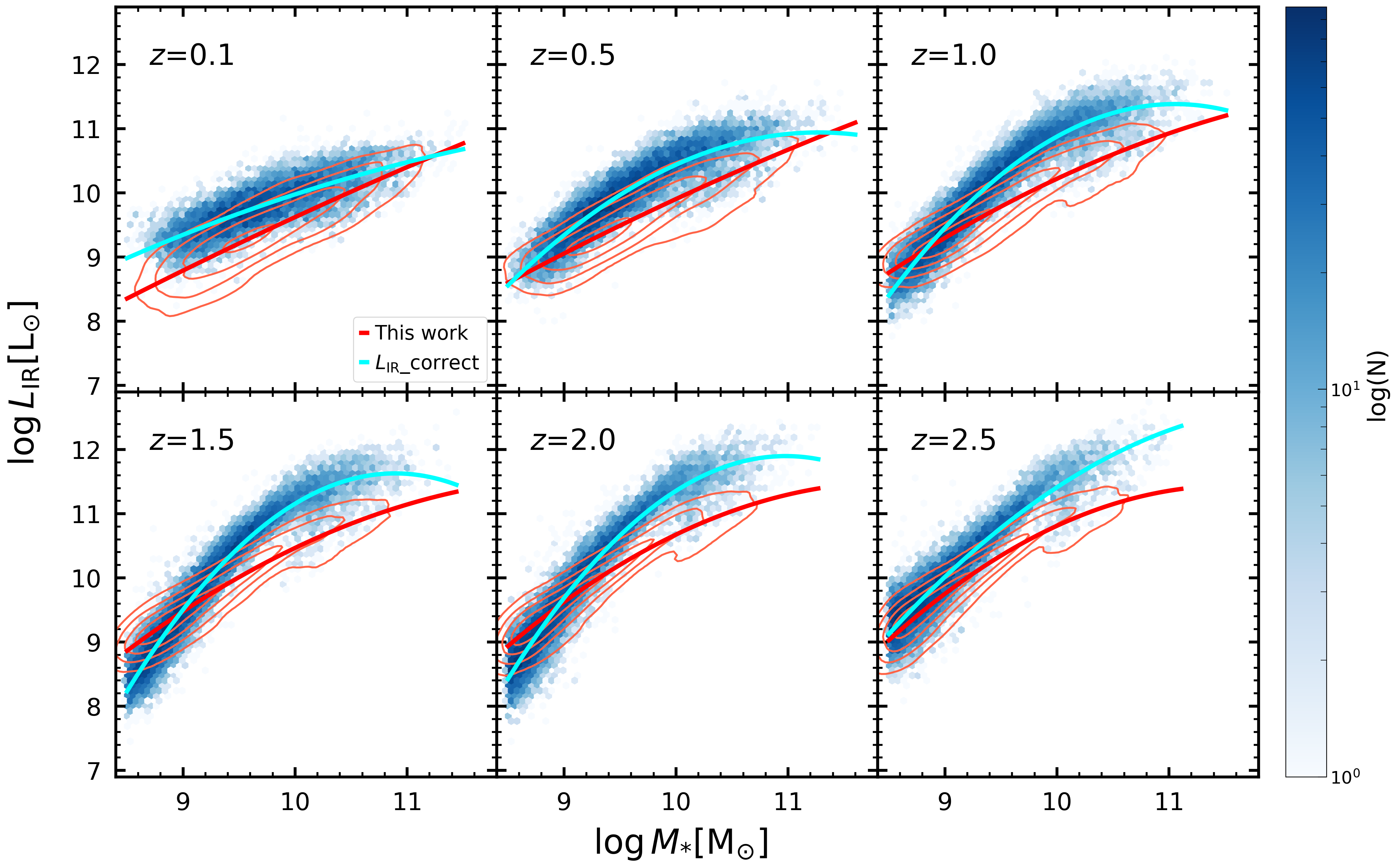}
    \caption{Comparison of the $M_{\ast}$--$L_{\rm IR}$ relations at six redshifts for the EAGLE SFGs with original (red contours) versus corrected (coloured hexagonal bins) $L_{\rm IR}$. The colour of the hexagonal bins represents the number of data points within that bin. The red and blue lines mark the best-fitting $M_{\ast}$--$L_{\rm IR}$ relations in the form of a second-order polynomial with original and corrected $L_{\rm IR}$.}\label{fig:figA4}
\end{figure*}

\begin{figure*}
	\centering
	\includegraphics[width=0.8\textwidth]{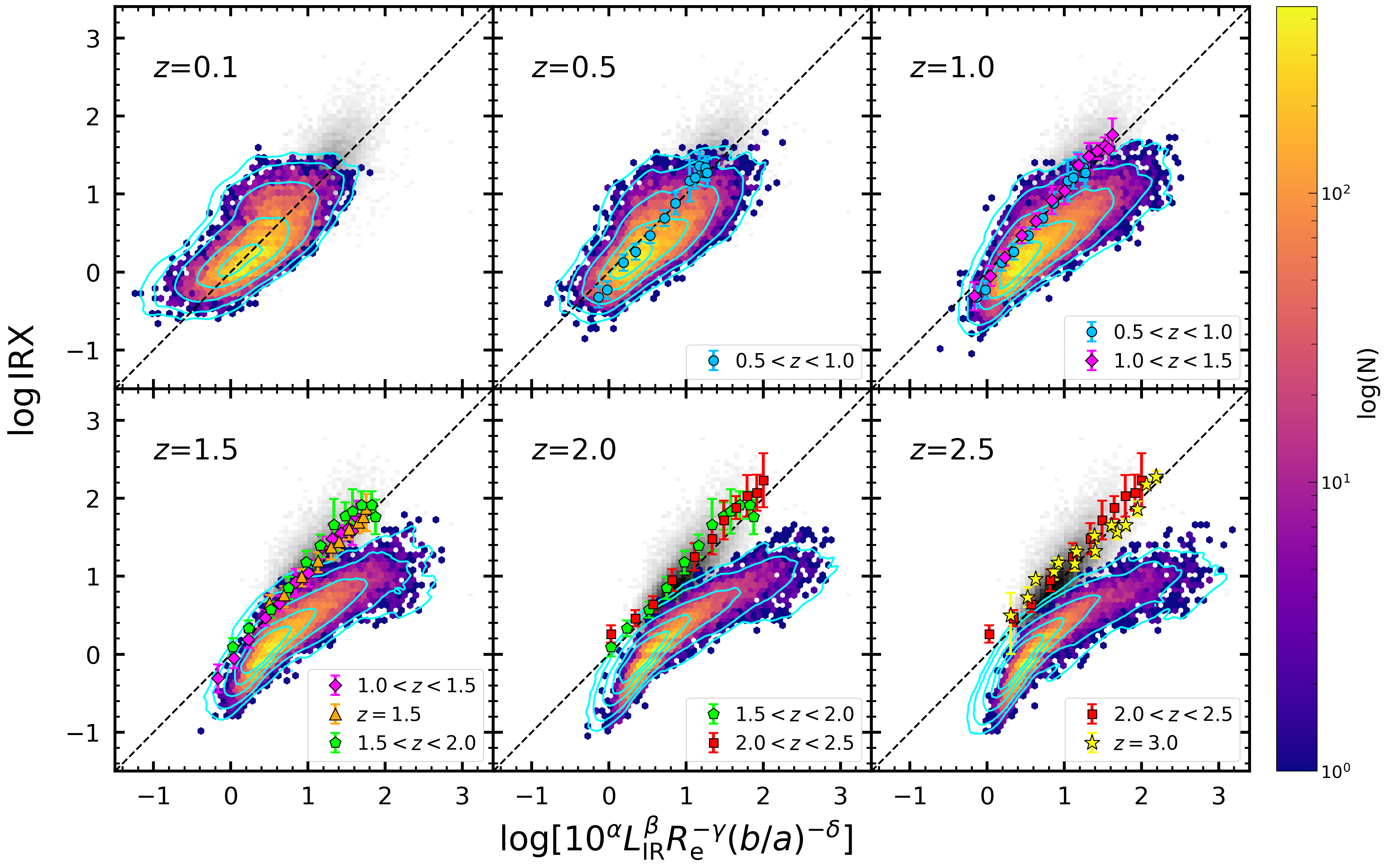}
    \caption{The predicted universal IRX relations for the EAGLE SFGs with corrected $R_{\rm e}$ at six redshifts. The gray-scale map and data points with error bars are the same as those in \protect Fig.~\ref{fig:fig2}. The colour of the hexagonal bins represents the number of SFGs within that bin. The cyan contours are the same as those in \protect Fig.~\ref{fig:fig5}. One can see that the locations of the majority of EAGLE SFGs with corrected $R_{\rm e}$ are only slightly changed compared to that with original $R_{\rm e}$.}\label{fig:figA5}
\end{figure*}

\begin{figure*}
	\centering
	\includegraphics[width=0.8\textwidth]{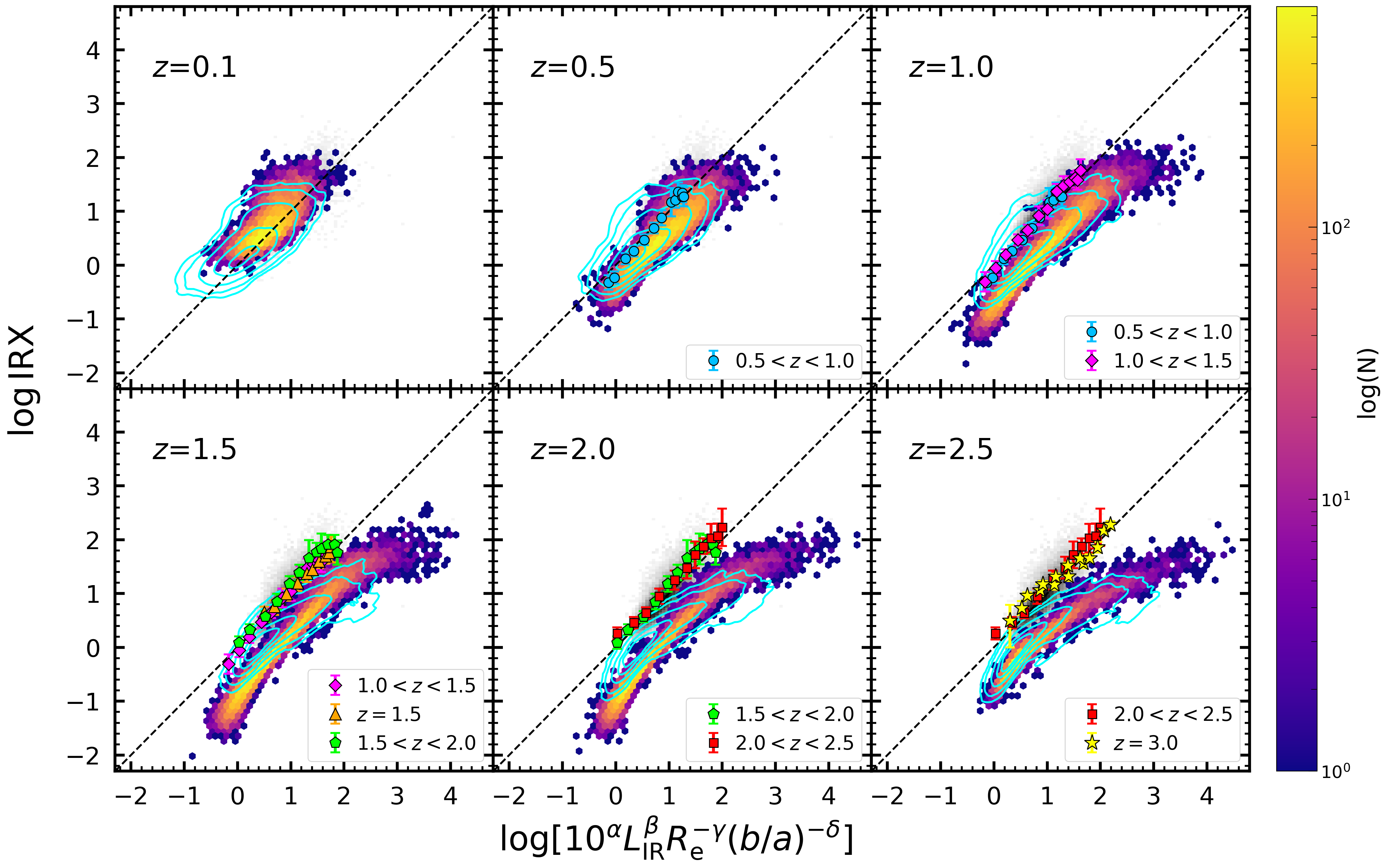}
    \caption{The predicted universal IRX relations for the EAGLE SFGs with corrected $L_{\rm IR}$ and IRX at six redshifts. The gray-scale map and data points with error bars are the same as those in \protect Fig.~\ref{fig:fig2}. The colour of the hexagonal bins represents the number of our EAGLE SFGs with corrected $L_{\rm IR}$ and IRX within that bin. The cyan contours are the same as those in \protect Fig.~\ref{fig:fig5}.}\label{fig:figA6}
\end{figure*}

\begin{figure*}
	\centering
	\includegraphics[width=0.95\textwidth]{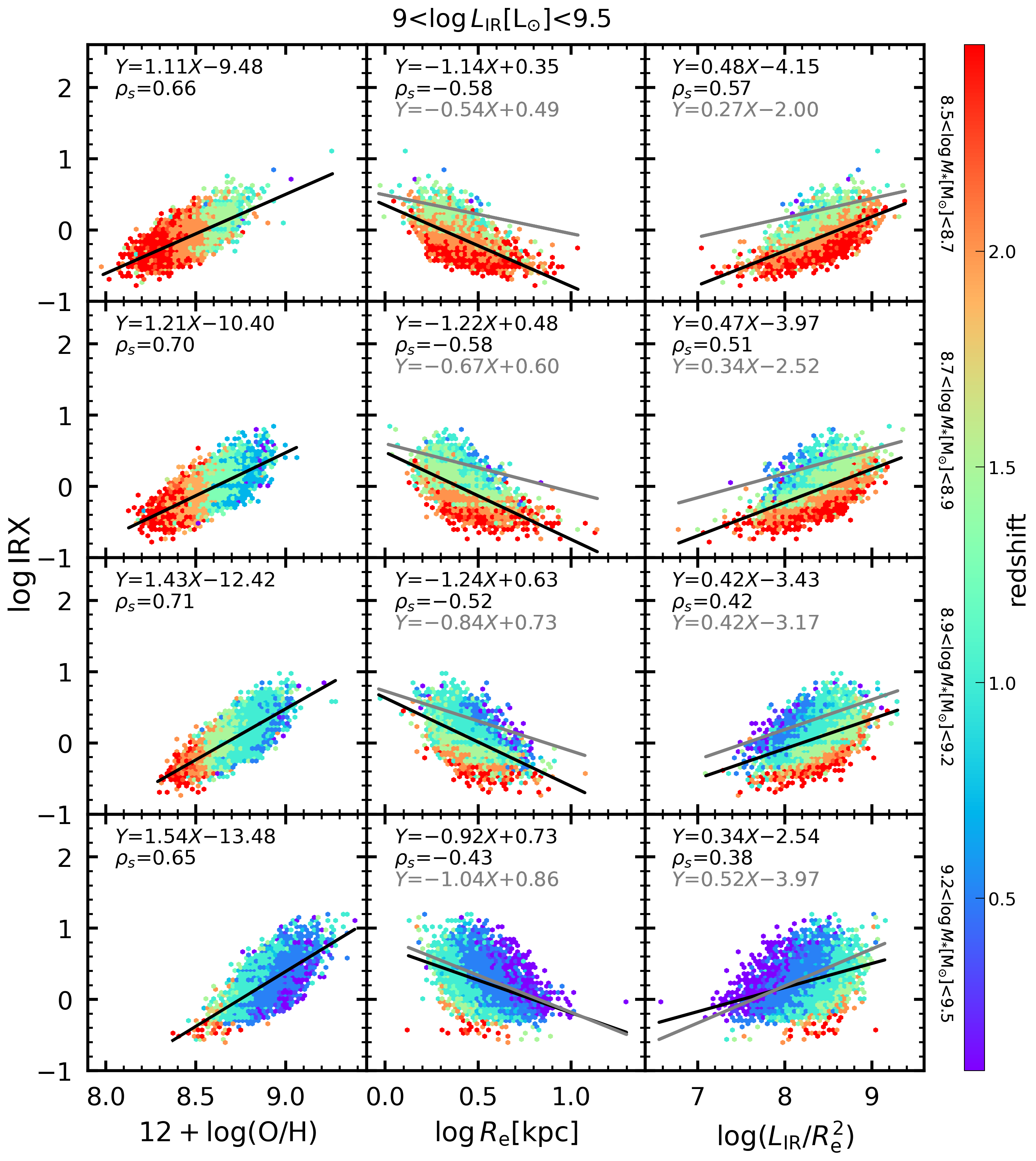}
	\caption{Relationships between IRX and $Z$, $R_{\rm e}$ and $L_{\rm IR}/R_{\rm e}^2$ (from left to right) for subsamples of the EAGLE SFGs with $9<\log (M_{\ast}/$M$_\odot)<9.5$ split into four stellar mass bins over $8.5<\log (M_{\ast}/$M$_\odot)<9.5$ (from top to bottom). Each hexagonal bin is colour coded by the median redshift of subsample SFGs within that bin. The black and grey solid lines denote the best-fitting power-law relations in this work and from \protect\citet{Qin2019}, and the relevant best-fitting functions are shown at the top of each panel. The coefficient of Spearman's rank $\rho_{\rm s}$ is presented to show the strength of the correlation.}\label{fig:figA7}
\end{figure*}

\begin{figure*}
	\centering
	\includegraphics[width=0.95\textwidth]{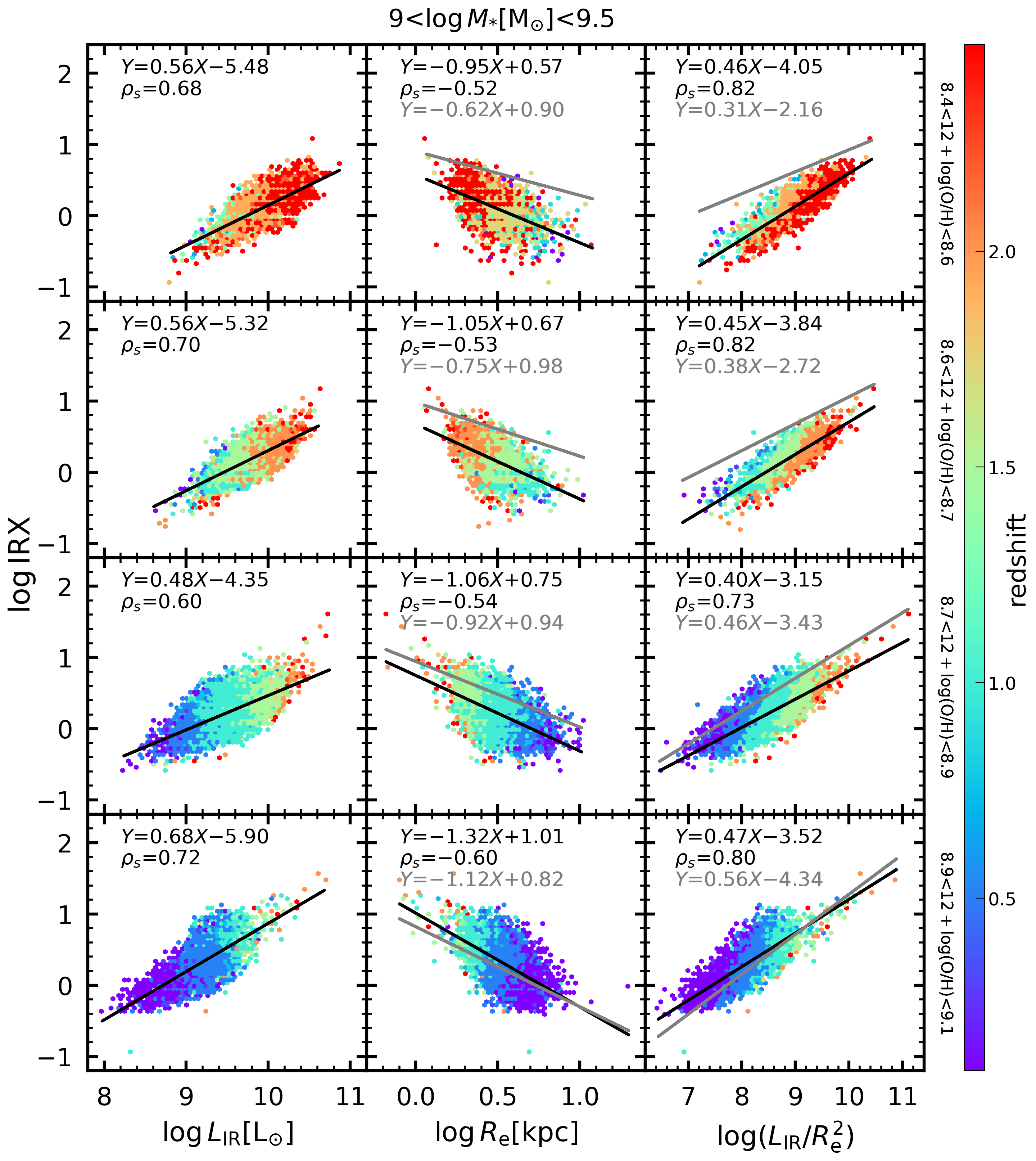}
	\caption{Relationships between IRX and $L_{\rm IR}$, $R_{\rm e}$ and $L_{\rm IR}/R_{\rm e}^2$ (from left to right) for subsamples of the EAGLE SFGs with $9<\log(M_{\ast}/{\rm M_\odot})<9.5$ split into four metallicity bins over $8.4<12+\log({\rm O/H})<9.1$ (from top to bottom). Each hexagonal bin is colour coded by the median redshift of subsample SFGs within that bin. The black and grey solid lines denote the best-fitting power-law relations in this work and from \protect\citet{Qin2019}, and the relevant best-fitting functions are shown at the top of each panel. The coefficient of Spearman's rank $\rho_{\rm s}$ is presented to show the strength of the correlation.}\label{fig:figA8}
\end{figure*}

Figures~\ref{fig:figA1} to~\ref{fig:figA8} present more results from our analysis to support our conclusions. Details can be found in the caption of each plot.

\bsp
\label{lastpage}
\end{document}